\documentclass[11pt]{article}
\usepackage{amssymb}
\usepackage{amsthm}
\usepackage{amsmath}
\usepackage{fullpage,hyperref,mathrsfs,txfonts,epsfig,graphicx,graphics}

\newcommand\remove[1]{}

\newcommand{\rnote}[1]{}
\newcommand{\jnote}[1]{}

\renewcommand{\L}{\mathscr{L}}

\newcommand{\1}{\mathbf{1}}
\newcommand{\e}{\varepsilon}
\newcommand{\R}{\mathbb{R}}
\newcommand{\E}{\mathbb{E}}

\newcommand{\SDP}{\mathrm{SDP}}

\renewcommand{\span}{\mathrm{\bf span}}
\newcommand{\sign}{\mathrm{\bf sign}}
\newcommand{\Clust}{\mathrm{\bf Clust}}

\def\calG{{\cal G}}

\def\calL{{\cal L}}
\def\calC{{\cal C}}

\def\calX{{\cal X}}
\def\calX{{\cal X}}

\def\chop{ {\rm chop}}
\newcommand{\eps}{\varepsilon}
\newcommand{\infl}{{\rm Inf}}

\newtheorem{theorem}{Theorem}[section]
\newtheorem{lemma}[theorem]{Lemma}

\newtheorem{corollary}[theorem]{Corollary}

\newtheorem{remark}{Remark}[section]

\date{}

\title{Approximate kernel clustering}

\author{Subhash Khot\footnote{Research supported in part by NSF CARREER award CCF-0643626,
and a Microsoft New Faculty Fellowship. }\\ Courant Institute of
Mathematical Sciences\\
{\tt khot@cims.nyu.edu} \and Assaf Naor\footnote{Research supported
by NSF grants CCF-0635078 and DMS-0528387.}\\ Courant Institute of
Mathematical Sciences\\{\tt naor@cims.nyu.edu}}
\begin{document}

\maketitle

\begin{abstract}
In the kernel clustering problem we are given a large $n\times n$
positive semi-definite matrix $A=(a_{ij})$ with
$\sum_{i,j=1}^na_{ij}=0$ and a small $k\times k$ positive
semi-definite matrix $B=(b_{ij})$. The goal is to find a partition
$S_1,\ldots,S_k$ of $\{1,\ldots n\}$ which maximizes the quantity
$$
\sum_{i,j=1}^k \left(\sum_{(p,q)\in S_i\times
S_j}a_{pq}\right)b_{ij}.
$$
We study the computational complexity of this generic clustering
problem which originates in the theory of machine learning. We
design a constant factor polynomial time approximation algorithm for
this problem, answering a question posed by Song, Smola, Gretton and
Borgwardt. In some cases we manage to compute the sharp
approximation threshold for this problem assuming the Unique Games
Conjecture (UGC). In particular, when $B$ is the $3\times 3$
identity matrix the UGC hardness threshold of this problem is
exactly $\frac{16\pi}{27}$. We present and study a  geometric
conjecture of independent interest which we show would imply that
the UGC threshold when $B$ is the $k\times k$ identity matrix is
$\frac{8\pi}{9}\left(1-\frac{1}{k}\right)$ for every $k\ge 3$.
\end{abstract}

\section{Introduction}

This paper is devoted to an investigation of the polynomial time
approximability of a generic clustering problem which originates in
the theory of machine learning. In doing so, we uncover a connection
with a continuous geometric/analytic problem which is of independent
interest. In~\cite{SSGB07} Song, Smola, Gretton and Borgwardt
introduced the following framework for {\em kernel clustering
problems}. Assume that we are given a centered kernel, i.e. an
$n\times n$ positive semidefinite matrix $A=(a_{ij})$ with real
entries such that $\sum_{i,j=1}^n a_{ij}=0$ (the assumption that the
kernel is centered is a commonly used normalization in learning
theory---see~\cite{SS01} for more information on this topic). Such
matrices arise, for example, as correlation matrices of random
variables $(X_1,\ldots,X_n)$ that measure attributes of certain
empirical data, i.e. $a_{ij}=\E\left[ X_iX_j\right]$.
We think of $n$ as very large, and our goal is
to ``cluster" the matrix $A$ to a much smaller $k\times k$ matrix in
such a way that certain features could still be extracted from the
clustered matrix. Formally, given a partition of $\{1,\ldots,n\}$
into $k$ sets $S_1,\ldots,S_k$, define the clustering of $A$ with
respect to this partition to be the $k\times k$ matrix, whose
$(i,j)^{\rm th}$ entry is
\begin{eqnarray}\label{eq:def clustering}
\sum_{(p,q)\in S_i\times S_j}a_{pq}.
\end{eqnarray}
Let $A(S_1,\ldots,S_k)$ denote the $k\times k$ matrix given
by~\eqref{eq:def clustering}. In the kernel clustering problem, we
are given a positive semidefinite $k\times k$ matrix $B=(b_{ij})$,
and we wish to find the clustering $A(S_1,\ldots,S_k)=C=(c_{ij})$ of
$A$, which is most similar to $B$ in the sense that $\sum_{i,j=1}^k
c_{ij}b_{ij}$, i.e its scalar product with $B$, is as large as
possible. In other words, our goal is to compute the number (and the
corresponding partition):
\begin{eqnarray}\label{eq:defcost}\Clust(A|B)&\coloneqq& \nonumber
\max\left\{\sum_{i,j=1}^k \left(\sum_{(p,q)\in S_i\times
S_j}a_{pq}\right)b_{ij}:\ \{S_1,\ldots,S_k\}\ \mathrm{is\ a\
partition\ of} \ \{1,\ldots,n\}\right\}\\
&=& \max\left\{\sum_{i,j=1}^k A(S_1,\ldots,S_k)_{ij}\cdot b_{ij}:\
\{S_1,\ldots,S_k\}\ \mathrm{is\ a\ partition\ of} \
\{1,\ldots,n\}\right\}\nonumber\\
&=& \max \left\{ \sum_{i,j=1}^n a_{ij} b_{\sigma(i) \sigma(j)}:\
\sigma:\{1,\ldots,n\}\to \{1,\ldots,k\}\right\}.
\end{eqnarray}

The flexibility in the above formulation of the kernel clustering
problem is clearly in the choice of comparison matrix $B$, which
allows us to enforce a wide-range of clustering criteria. Using the
statistical interpretation of $(a_{ij})$ as a correlation matrix, we
can think of the matrix $B$ as encoding our belief/hypothesis that
the empirical data has a certain structure, and the kernel
clustering problem aims to efficiently expose this structure.

Several explicit examples of useful ``test matrices" $B$ are
discussed in~\cite{SSGB07}, including hierarchical clustering and
clustering data on certain manifolds. We refer to~\cite{SSGB07} for
additional information which illustrates the versatility of this
general clustering problem, including its relation to the Hilbert
Schmidt Independence Criterion (HSIC) and various experimental
results. In~\cite{SSGB07} it was asked if there is a polynomial time
approximation algorithm for computing $\Clust(A|B)$. Here we obtain
a constant factor approximation algorithm for this problem, and
prove some computational hardness of approximation results.

Before stating our results in full generality we shall now present a
few simple illustrative examples. If $B=I_k$ is the $k\times k$
identity matrix, then thinking once more of $a_{ij}$ as correlations
$\E\left[X_iX_j\right]$, our goal is to find a partition
$S_1,\ldots,S_k$ of $\{1,\ldots,n\}$ which maximizes the quantity
$$
\sum_{i=1}^k \sum_{p,q\in S_i} \E \left[X_pX_q\right],
$$
i.e. we wish to cluster the variables so as to maximize the total
intra-cluster correlations. As we shall see below, our results yield
a polynomial time algorithm which approximates $\Clust(A|I_k)$ up to
a factor of $\frac{8\pi}{9}\left(1-\frac{1}{k}\right)$. In
particular, when $k=3$ we obtain a $\frac{16\pi}{27}$ approximation
algorithm, and we show that assuming the Unique Games Conjecture
(UGC) no polynomial time algorithm can achieve an approximation
guarantee which is smaller than $\frac{16\pi}{27}$. The Unique Games
Conjecture was posed by Khot in~\cite{Khot02}, and it will be
described momentarily. For the readers who are not familiar with
this computational hypothesis and its remarkable applications to
hardness of approximation, it suffices to say that this hardness
result should be viewed as strong evidence that $\frac{16\pi}{27}$
is the sharp threshold below which no polynomial time algorithm can
solve the kernel clustering problem when $B=I_3$. Moreover, we
conjecture that $\frac{8\pi}{9}\left(1-\frac{1}{k}\right)$ is the
sharp approximability threshold (assuming UGC) for $\Clust(A|I_k)$
for every $k\ge 3$. In this paper, we reduce this conjecture  to a
purely geometric/analytic conjecture, which we will describe in
detail later, and prove some partial results about it.

Another illustrative example of the kernel clustering problem is the
case
\begin{equation*}
 B=   \begin{pmatrix}
  1 & -1 \\
   -1 & 1
   \end{pmatrix}.
   \end{equation*}
In this case, we clearly have
\begin{eqnarray}\label{eq:grothendieck}
\Clust\left(A\left|\begin{pmatrix}
  1 & -1 \\
   -1 & 1
   \end{pmatrix}\right.\right)=\max\left\{\sum_{i,j=1}^n: a_{ij}\e_i\e_j:\ \e_1,\ldots,\e_n\in
   \{-1,1\}\right\}.
\end{eqnarray}
The optimization problem in~\eqref{eq:grothendieck} is well known as
the positive semi-definite Grothendieck problem and has several
algorithmic applications (see~\cite{Rietz74,Nes98,AN06,CMM07}). It
has been shown by Rietz~\cite{Rietz74} that the natural semidefinite
relaxation of~\eqref{eq:grothendieck} has integrality gap
$\frac{\pi}{2}$ (see also Nesterov's work~\cite{Nes98}). Our results
imply that assuming the UGC $\frac{\pi}{2}$ is the sharp
approximation threshold for the positive-semidefinite Grothendieck
problem. Note that without the assumption that $A$ is positive
semidefinite the natural semidefinite relaxation
of~\eqref{eq:grothendieck} has integrality gap $\Theta(\log n)$.
See~\cite{NRT99,CW04,AMMN06} for more information,
and~\cite{ABKHS05} for hardness results for this problem.

We can also view the problem~\eqref{eq:grothendieck} as a
generalization of the MaxCut problem. Indeed, let
$G=(V=\{1,\ldots,n\},E)$ be an $n$-vertex loop-free graph. For every
vertex $i\in V$ let $d_i$ denote its degree in $G$. Let $A$ be the
Laplacian of $G$, i.e. $A$ is the $n\times n$ matrix given by
\begin{eqnarray}\label{eq:laplacian}
a_{ij}=\left\{\begin{array}{ll} d_i & \mathrm{if}\ i=j,\\-1 &
\mathrm{if}\ i\neq j\ \wedge\ ij\in E,\\
0 & \mathrm{if}\ i\neq j\ \wedge\ ij\notin E.
\end{array}\right.
\end{eqnarray}
Then $A$ is positive semi-definite since it is diagonally dominant.
For every $\e_1,\ldots,\e_n\in \{-1,1\}$ let $S\subseteq V$ be the
set $S\coloneqq \{i\in V:\ \e_i=1\}$. Then:
\begin{multline}\label{eq:maxcut}
\sum_{i,j=1}^n a_{ij} \e_i\e_j= \sum_{i=1}^n d_i -2|E(S,S)| -
2|E(V\setminus S,V\setminus S)|+2|E(S,V\setminus
S)|\\=2|E|-2\left(|E|-|E(S,V\setminus S)|\right)+2|E(S,V\setminus
S)|=4|E(S,V\setminus S)|.
\end{multline}
Hence
$$
\Clust\left(A\left|\begin{pmatrix}
  1 & -1 \\
   -1 & 1
   \end{pmatrix}\right.\right)=4\mathrm{MaxCut}(G).
$$
Using H{\aa}stad's inapproximability result for MaxCut~\cite{Haa01}
it follows that if $P\neq NP$ there is no polynomial time algorithm
which approximates~\eqref{eq:grothendieck} up to a factor smaller
than $\frac{17}{16}$.

\medskip

\noindent{\bf Our algorithmic results.} For a fixed positive
semidefinite matrix $B$, the approximability threshold for the
problem of computing $\Clust(A|B)$ depends on $B$.  It is therefore
of interest to study the performance of our algorithms in terms of
the matrix $B$. We do obtain
 bounds which depend on $B$ (which are probably suboptimal in general)---the precise statements are contained in
 Theorem~\ref{thm:analysis} and Theorem~\ref{thm:noncentered}. For
 the sake of simplicity, in the introduction we state bounds which
 are independent of $B$. We believe that the problem of computing
the approximation threshold (perhaps under UGC) for each fixed $B$
is an interesting problem which deserves further research.

If $A$ is centered, i.e. $\sum_{i,j=1}^na_{ij}=0$, then for every
$k\times k$ positive semi-definite matrix $B$ our algorithm achieves
an approximation ratio of $\pi\left(1-\frac{1}{k}\right)$. If, in
addition, $B$ is centered and spherical, i.e.
$\sum_{i,j=1}^kb_{ij}=0$ and $b_{ii}=1$ for all $i$, then our
algorithm achieves an approximation ratio of
$\frac{8\pi}{9}\left(1-\frac{1}{k}\right)$. This ratio is also valid
if $B$ is the identity matrix, and as we mentioned above, we believe
that this approximation guarantee cannot be improved assuming the
UGC (and here we prove this conjecture for $k=3$). When $A$ is not
necessarily centered (note that this case is of lesser interest in
terms of the applications in machine learning) we obtain an
algorithm which achieves an approximation ratio of
$1+\frac{3\pi}{2}$ (this is probably sub-optimal). All of our
algorithms, which are described in Section~\ref{sec:alg}, use
semi-definite programming in a perhaps non-obvious way. The rounding
algorithm of our semi-definite relaxation amounts to proving certain
geometric inequalities which can be viewed as variants of the
positive semi-definite Grothendieck inequality. This analysis is
presented in Section~\ref{sec:alg}. As a concrete example we state
in this introduction the following Grothendieck-type inequality
which corresponds to our $\frac{8\pi}{9}\left(1-\frac{1}{k}\right)$
algorithm:
\begin{theorem}\label{thm:new-grothendieck} Let $(a_{ij})$ be an $n\times n$ positive semi-definite matrix with
$\sum_{i,j=1}^n a_{ij}=0$. Then for every $k\ge 3$ and
$v_1,\ldots,v_k\in S^{k-1}$ we have
\begin{equation}\label{eq:our Grothendieck}
\max_{x_1,\ldots,x_n\in S^{n-1}}\sum_{i,j=1}^n a_{ij} \langle
x_i,x_j\rangle \le
\frac{8\pi}{9}\left(1-\frac{1}{k}\right)\max_{\sigma:\{1,\ldots,n\}\to
\{1,\ldots,k\}}\sum_{i,j=1}^na_{ij}\langle
v_{\sigma(i)},v_{\sigma(j)}\rangle.
\end{equation}
\end{theorem}
Inequality~\eqref{eq:our Grothendieck} is sharp when $k=3$, and we
conjecture that it is sharp for all $k\ge 4$. This conjecture is
related to a geometric conjecture which we describe below.

\medskip

\noindent{\bf The Unique Games Conjecture, hardness of
approximation, and the propeller problem.} Our hardness result for
kernel clustering problem is based on the Unique Games Conjecture
which was put forth by Khot in~\cite{Khot02}. We shall now describe
this conjecture. A {\em Unique Game} is an optimization problem with
an instance ${\L}= {\L}( G(V,W,E), n, \{\pi_{vw}\}_{(v,w) \in W})$.
Here $G(V,W,E)$ is a regular bipartite graph with vertex sets $V$
and $W$ and edge set $E$. Each vertex is supposed to receive a label
from the set $\{1,\ldots, n\}$.
For every edge $(v,w) \in E$ with $v \in V$ and $w\in W$, there is a
given permutation $\pi_{vw}: \{1,\ldots,n\} \to \{1,\ldots, n\}$. A
labeling to the Unique Game instance is an assignment $ \rho: V \cup
W \to \{1,\ldots, n\}$. An edge $(v,w)$ is satisfied by a labeling
$\rho$ if and only if $\rho(v) = \pi_{vw}(\rho(w))$. The goal is to
find a labeling that maximizes the fraction of edges satisfied (call
this maximum ${\rm OPT}({\L}))$. We think of the number of labels
$n$ as a constant and the size of the graph $G(V,W,E)$ as the size
of the problem instance.

The Unique Games Conjecture asserts that for  arbitrarily small
constants $\e, \delta > 0$, there exists a constant $n = n(\e,
\delta)$ such that no polynomial time algorithm can distinguish
whether a Unique Games instance ${\L}$ with $n$ labels satisfies
${\rm OPT}({\L}) \geq 1-\eps$ or ${\rm OPT}({\L}) \leq
\delta$\footnote{As stated in~\cite{Khot02}, the conjecture says
that it is NP-hard to distinguish between these two cases. However
if one only wants to rule out polynomial time algorithms, the
conjecture as stated here suffices.}. This conjecture  is (by now) a
commonly used complexity assumption to prove hardness of
approximation results. Despite several recent attempts to get better
polynomial time approximation algorithms  for the Unique Game
problem (see the table in~\cite{CMM06} for a description of known
results), the unique games conjecture still stands.

 Our UGC
hardness result for kernel clustering, which is presented in
Section~\ref{sec:UGC}, is based at heart on the ``dictatorship vs.
low-influence"
 paradigm that is recurrent in UGC hardness results (for
example~\cite{Khot02,KKMO07}). In order to apply this paradigm one
usually designs a probabilistic test on a given Boolean function on
the Boolean hypercube and then analyzes the acceptance probability
of this test in the two extremes of dictatorship functions and
functions without influential variables. The gap between these two
acceptance probabilities translates into the hardness of
approximation factor. In our case, instead of a probabilistic test
we need to design a positive semidefinite quadratic form on the
truth table of the function. Our form is the sum of the squares of
the Fourier coefficients of level $1$. This already yields
$\frac{\pi}{2}$ UGC hardness when $k=2$. For larger $k$ we need to
work with functions from $\{1,\ldots,k\}^n$ to $\{1,\ldots,k\}$. The
analysis of this approach leads to the  ``propeller problem" which
we now describe. The details of this connection are explained in
Section~\ref{sec:UGC}.

We believe that one of the interesting aspects of the present paper
is that complexity considerations lead to geometric/analytic
problems which are of independent interest. Similar such connections
have been recently discovered in~\cite{KKMO04,FKO07}. In our case
the reduction from UGC to kernel clusterings leads to the following
question, which we call the ``propeller problem" for reasons that
will become clear presently. Let $\gamma_{k-1}$ denote the standard
Gaussian measure on $\R^{k-1}$, i.e. the density of $\gamma_{k-1}$
is $(2\pi)^{-(k-1)/2}e^{-\|x\|_2^2/2}$. Let $A_1,\ldots,A_k$ be a
partition of $\R^{k-1}$ into measurable sets. For each $i\in
\{1,\ldots,k\}$ consider the Gaussian moment of the set $A_i$, i.e.
the vector
$$
z_i\coloneqq \int_{A_i} xd\gamma_{k-1}(x)\in \R^{k-1}.
$$
Our goal is to find the partition which maximizes the sum of the
squared Euclidean lengths of the Gaussian moments of the elements of
the partition, i.e. $ \sum_{i=1}^k \|z_i\|_2^2$. Let $C(k)$ denote
the value of this maximum (in Section~{\ref{sec:geometry} we show
that this is indeed a maximum and not just a supremum). In
Section~\ref{sec:UGC} we show that assuming the UGC there is no
polynomial time algorithm which approximates $\Clust(A|I_k)$ to a
factor smaller than $\frac{1-1/k}{C(k)}$. In
Section~\ref{sec:geometry} we show that $C(2)=\frac{2}{\pi}$ and
$C(3)=\frac{9}{8\pi}$. The value of $C(3)$ comes from the partition
of the plane $\R^2$ into a ``propeller", i.e. three cones of angle
$\frac{2\pi}{3}$ with cusp at the origin. Most of
Section~\ref{sec:geometry} is devoted to the proof of the following
theorem:
\begin{theorem}\label{thm:reduction to simplicial}
$C(k)$ is attained at a {\bf simplicial conical partition}, i.e. a
partition $A_1,\ldots, A_k$ of $\R^{k-1}$ which has the following
form: let $A_1,\ldots,A_m$ be the elements of the partition which
have positive measure. Then $A_j=B_j\times \R^{k-m}$ where
$B_j\subseteq \R^{m-1}$ is a cone with cusp at $0$ whose base is a
simplex.
\end{theorem}
It is tempting to believe that the optimal simplicial conical
partition described in Theorem~\ref{thm:reduction to simplicial}
occurs when the cones $B_1,\ldots, B_m$ are generated by the regular
simplex. However, in Section~\ref{sec:geometry} we prove that among
such {\bf regular simplicial conical partitions} the one which
maximizes the sum of the squared lengths of its Gaussian moments is
when $m=3$.

We therefore conjecture that for every $k\ge 3$ an optimal partition
for the problem described above
 is actually $\{C_1\times \R^{k-3}, C_2\times
\R^{k-3}, C_3\times \R^{k-3}\}$, where $\{C_1,C_2,C_3\}$ is the
propeller partition of $\R^2$---see Figure~\ref{fig:manor}. If this
``propeller conjecture" holds true then it would follow that our
$\frac{8\pi}{9}\left(1-\frac{1}{k}\right)$ approximation algorithm
is optimal assuming the UGC for every $k\ge 4$, and not just for
$k\in \{2,3\}$. The full propeller conjecture seems to be a
challenging geometric problem of independent interest, not just due
to the connection that we establish between it and the study of
hardness of approximation for kernel clustering.


\begin{figure}[h]\label{fig:manor}
\begin{center}\includegraphics[scale=2.7]{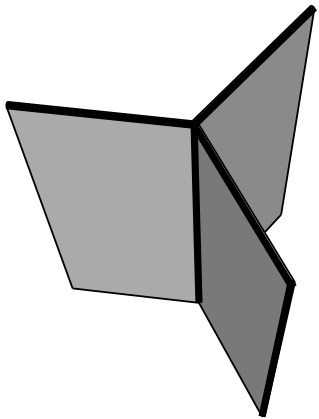}
\end{center}
 \caption{{{\em  The conjectured optimal
 partition for the ``sum of squares of Gaussian moments problem" described above consists of a partition of $R^{k-1}$
 into $3$ parts, and the remaining $k-3$ parts are empty. This partition corresponds to a planar $120^\circ$
 ``propeller" multiplied by an orthogonal copy of $\R^{k-3}$. }}} \label{fig:midpoints}
\end{figure}

\bigskip


We end this introduction with an explanation of how our work relates
to the recent result of Raghavendra~\cite{Prasad} which shows that
for any generalized constraint satisfaction problem\footnote{In a
generalized CSP, every assignment to variables in a constraint has a
real-valued (possibly negative) pay-off instead of a simple decision
saying that the assignment is a satisfying assignment or not.} (CSP)
there is a generic way of writing a semidefinate relaxation that
achieves an optimal approximation ratio assuming the Unique Games
Conjecture. Our clustering problem fits in the framework
of~\cite{Prasad} as follows: we wish to compute
\begin{equation} \max \left\{ \sum_{i,j=1}^n
a_{ij} b_{\sigma(i) \sigma(j)}:\ \sigma:\{1,\ldots,n\}\to
\{1,\ldots,k\}\right\}, \label{eqn:clustering-problem}
\end{equation}
where $(a_{ij})$ is a centered positive semi-definite matrix and
$(b_{ij})$ is a positive semi-definite matrix. One can think of this
problem as a CSP (with an extra global constraint corresponding to
the positive semi-definiteness) where the set of variables is
$\{1,\ldots,n\}$ and we wish to assign each variable a value from
the domain $\{1,\ldots,k\}$. For every pair $(i,j) \in
\{1,\ldots,n\} \times \{1,\ldots,n\}$, there is a constraint with
weight $a_{ij}$. We get a payoff of $b_{st}$ if variables $i$ and
$j$ are assigned $s \in \{1,\ldots,k\}$ and $t \in \{1,\ldots,k\}$
respectively.

Raghavendra shows that every integrality gap instance for his
generic SDP relaxation can be translated into a UGC-hardness of
approximation result with the hardness factor (essentially) the same
as the integrality gap. We make here the non-trivial observation
 that in the reduction of~\cite{Prasad},
starting with an integrality gap instance for (the generic SDP
relaxation of) the clustering problem
(\ref{eqn:clustering-problem}),  the matrix of the constraint
weights $(a_{ij})$  indeed turns out to be positive semi-definite as
required in the kernel clustering problem (this requires proof---the
details are omitted  since this is a digression from the topic of
this paper). Thus Raghavendra's result can be made to apply to the
kernel clustering problem (i.e. the generic SDP achieves the optimum
approximation ratio assuming UGC).

Nevertheless, it is also useful to look at different relaxations and
rounding procedures for the following reasons. Firstly, for a given
problem there could be an SDP relaxation that is more {\it natural}
than the generic one and might be easier to work with. Secondly,
Raghavendra's result (that the integrality gap is same as the
hardness factor) applies only when the integrality gap is a
constant. This is a priori not clear for the kernel clustering
problem. For instance, a priori the integrality gap could be
$\Omega(\log n)$ (as is the case for Grothendieck problem on a
general graph---see~\cite{AMMN06}). So before applying the result
of~\cite{Prasad}, one would need to show that the integrality gap of
the generic SDP is indeed a constant. Thirdly, for CSPs with
negative payoffs (as is the case in the kernel clustering problem),
Raghavendra shows that the {\it value} computed by the generic SDP
achieves the optimal approximation ratio (modulo UGC), but the paper
does not give a rounding procedure. Finally, Raghavendra's result
does not really shed light on the exact hardness threshold in the
sense that it shows how to translate integrality gap instances into
a UGC hardness result, but gives no idea as to how to construct an
integrality gap instance in the first place. Constructing the
integrality gap instance in general amounts to answering certain
isoperimetric type geometric question (naturally leading to a
dictatorship test, or the other way round. In other words, the
geometric question itself might be inspired by the dictatorship test
that we have in mind). Thus as far as we know, we cannot avoid
designing an explicit dictatorship test and answering an
isoperimetric type question, whether or not we start with
Raghavendra's generic SDP that is guaranteed to be optimal. As
mentioned before, in the clustering problem where $B=(b_{st})$ is
centered and spherical, we show that the UGC-hardness threshold is
at least $\frac{1-1/k}{C(k)}$ and characterizing $C(k)$ seems to be
a challenging geometric question.

\section{Constant factor approximation algorithms for kernel
clustering}\label{sec:alg}

Let $A\in M_n(\R)$ and $B\in M_k(\R)$ be positive semidefinite
matrices. Then there are $u_1,\ldots,u_n\in \R^n$ and
$v_1,\ldots,v_k\in \R^k$ such that $a_{ij}=\langle u_i,u_j\rangle$
and $b_{ij}=\langle v_i,v_j\rangle$. Such vectors can be found in
polynomial time (this is simply the Cholesky decomposition). The
instance of the kernel clustering problem will be called centered if
$\sum_{i,j=1}^n a_{ij}=0 $, or equivalently $\sum_{i=1}^n u_i=0$.
The instance will be called spherical if $b_{ii}=1=\|v_i\|_2^2$ for
all $i\in \{1,\ldots,k\}$. Let $R(B)$ be the radius of the smallest
Euclidean ball containing $\{v_1,\ldots,v_k\}$. Note that $R(B)$ is
indeed only a function of $B$, i.e. it does not depend on the
particular representation of $B$ as a Gram matrix. Moreover, it is
possible to compute $R(B)$, and given the decomposition
$b_{ij}=\langle v_i,v_j\rangle$ a vector $w\in \R^k$ such that
$\max_{j\in \{1,\ldots,k\}}\|v_j-w\|_2=R(B)$, in polynomial time
(see~\cite{GK92}).

Our goal is to compute in polynomial time the quantity:
$$
\Clust(A|B)\coloneqq \max_{\sigma:\{1,\ldots,n\}\to
\{1,\ldots,k\}}\sum_{i,j=1}^n
a_{ij}b_{\sigma(i)\sigma(j)}=\max_{\sigma:\{1,\ldots,n\}\to
\{1,\ldots,k\}}\sum_{i,j=1}^n\langle u_i,u_j\rangle\langle
v_{\sigma(i)},v_{\sigma(j)}\rangle.
$$
Our algorithm, which is based on semidefinite programming, proceeds
via the following steps:
\begin{enumerate}
\item Compute a Cholesky decomposition of $B$, i.e. $v_1,\ldots, v_k\in \R^{k}$ with $b_{ij}=\langle v_i,v_j
\rangle$.
\item Compute (using for example~\cite{GK92}) $R(B)$ and a vector $w\in \R^{k}$ such
that
$$\max_{j\in \{1,\ldots,k\}}\|v_j-w\|_2=R(B).
$$
\item Solve the semidefinite program \begin{eqnarray*}\label{eq:thSDP}\!\!\!\!\!\max\left\{
  \sum_{i,j=1}^n a_{ij} \cdot  \left\langle
   \|w\|_2 u + R(B) x_i, \ \|w\|_2 u + R(B) x_j \right\rangle :\ u,x_1,\ldots,x_n\in \R^{n+1}
   \ \wedge\ \|u\|_2=1\ \wedge\  \forall i\  \|x_i\|_2\le
   1\right\}.
\end{eqnarray*}
\item Choose $p,q\in \{1,\ldots,k\}$ such that
$\|v_p-v_q\|_2=\max_{i,j\in \{1,\ldots,k\}}\|v_i-v_j\|_2$. Let
$g_1,g_2\in \R^{n+1}$ be i.i.d. standard Gaussian vectors and define
$\sigma: \{1,\ldots,n\}\to \{1,\ldots,k\}$ by
\begin{eqnarray}\label{eq:def sigma}
\sigma(r)=\left\{\begin{array}{ll} p & \mathrm{if}\ \langle
g_1,x_r\rangle \ge \langle g_2,x_r\rangle,\\
q & \mathrm{if}\ \langle g_2,x_r\rangle \ge \langle g_1,x_r\rangle.
\end{array}\right.
\end{eqnarray}
\item Choose distinct $\alpha,\beta,\gamma\in \{1,\ldots,k\}$ such that
$$
\left\|v_\alpha-\frac{v_\alpha+v_\beta+v_\gamma}{3}\right\|_2^2+\left\|v_\beta-
\frac{v_\alpha+v_\beta+v_\gamma}{3}\right\|_2^2+\left\|v_\gamma-\frac{v_\alpha+v_\beta+v_\gamma}{3}\right\|_2^2
$$
is maximized among all such choices of $\alpha,\beta,\gamma$. Let
$g_1,g_2,g_3\in \R^{n+1}$ be i.i.d. standard Gaussian vectors and
define $\tau:\{1,\ldots,n\}\to \{1,\ldots,k\}$ by
\begin{eqnarray}\label{eq:def tau}
\tau(r)=\left\{\begin{array}{ll} \alpha & \mathrm{if}\ \langle
g_1,x_r\rangle \ge \max
\left\{\langle g_2,x_r\rangle,\langle g_3,x_r\rangle\right\},\\
\beta & \mathrm{if}\ \langle g_2,x_r\rangle \ge \max \left\{\langle
g_1,x_r\rangle,\langle g_3,x_r\rangle\right\}, \\
\gamma & \mathrm{if}\ \langle g_3,x_r\rangle \ge \max \left\{\langle
g_1,x_r\rangle,\langle g_2,x_r\rangle\right\}.
\end{array}\right.
\end{eqnarray}
\item Output $\sigma$ if $\sum_{i,j=1}^n
a_{ij}b_{\sigma(i)\sigma(j)}\ge \sum_{i,j=1}^n
a_{ij}b_{\tau(i)\tau(j)}$. Otherwise output $\tau$.
\end{enumerate}

\begin{remark}\label{rem:s}{\em
The astute reader might notice that there is an obvious
generalization of the above algorithm. Namely for every fixed
integer $s\in [2,k]$ we can choose a subset $S\subseteq \{1,..,k\}$
of cardinality $s$ which maximizes the quantity
$$
\sum_{i\in S} \left\|v_i-\frac{1}{s}\sum_{j\in S} v_j\right\|_2^2.
$$
Then, we can choose $s$ i.i.d. standard Gaussians $\{g_i\}_{i\in
S}\subseteq \R^{n+1}$ and define $\sigma_s:\{1,\ldots,n\}\to
\{1,\ldots,k\}$ analogously to the above, namely $\sigma_s(r)=i$ if
$$
\langle g_i,x_r\rangle=\max_{j\in S} \langle g_j,x_r\rangle.
$$
Then, we can consider the assignments
$\sigma_2,\sigma_3,\ldots,\sigma_s$ and choose the one which
maximizes the objective $\sum_{i,j=1}^n
a_{ij}b_{\sigma_\ell(i)\sigma_\ell(j)}$. In spite of this
flexibility, it turns out that the the rounding method described
above does not improve if we take $s\ge 4$. In order to demonstrate
this fact we will proceed below to analyze the algorithm for general
$s$, and then optimize over $s$.}
\end{remark}

Bounds on the performance of the above algorithm are contained in
the following theorem:

\begin{theorem}\label{thm:analysis} Assume that $A$ is centered, i.e. that
$\sum_{i,j=1}^n a_{ij}=0$. Let $p,q,\alpha,\beta,\gamma\in
\{1,\ldots,k\}$ and $v_1,\ldots,v_k$ be as in the description above.
Then the algorithm outputs in polynomial time a random assignment
$\lambda:\{1,\ldots,n\}\to \{1,\ldots,k\}$ satisfying
\begin{multline}\label{eq:ugly}
\Clust(A|B)\\\le  \min\left\{\frac{2\pi
R(B)^2}{\|v_p-v_q\|_2^2},\frac{16\pi
R(B)^2}{9\left(\left\|v_\alpha-\frac{v_\alpha+v_\beta+v_\gamma}{3}\right\|_2^2+\left\|v_\beta-
\frac{v_\alpha+v_\beta+v_\gamma}{3}\right\|_2^2+
\left\|v_\gamma-\frac{v_\alpha+v_\beta+v_\gamma}{3}\right\|_2^2\right)}\right\}\E\left[\sum_{i,j=1}^n
a_{ij}b_{\lambda(i)\lambda(j)}\right].
\end{multline}
In particular we always have
\begin{equation}\label{eq:centered}
\Clust(A|B)\le \pi\left(1-\frac{1}{k}\right)\E\left[\sum_{i,j=1}^n
a_{ij}b_{\lambda(i)\lambda(j)}\right],
\end{equation}
and if $B$ is centered and spherical, i.e. $\sum_{i,j=1}^kb_{ij}=0$
and $b_{ii}=1$ for all $i$, then
\begin{equation}\label{eq:spherical}
\Clust(A|B)\le
\frac{8\pi}{9}\left(1-\frac{1}{k}\right)\E\left[\sum_{i,j=1}^n
a_{ij}b_{\lambda(i)\lambda(j)}\right].
\end{equation}
The same bound in~\eqref{eq:spherical} holds true if $B$ is the
identity matrix.
\end{theorem}
We single out in the next theorem the case $k\in \{2,3\}$, since in
these cases we have matching UGC hardness results. Note that for
general $k$ we obtain a factor $\pi$ approximation algorithm,
answering positively the question posed by Song, Smola, Gretton and
Borgwardt in~\cite{SSGB07}.

\begin{theorem}\label{thm:tight UGC} Assume that $A$ is centered and $B$ is a $2\times 2$
matrix. Then our algorithm achieves a $\frac{\pi}{2}$ approximation
factor. Assuming the Unique Games Conjecture no polynomial time
algorithm achieves an approximation guarantee smaller than
$\frac{\pi}{2}$ in this case.

Assume that $A$ is centered, $k=3$ and $B$ is centered and spherical
(since $k=3$ this forces $B$ to be the Gram matrix of the-degree
three roots of unity in the complex plane). Then our algorithm
achieves an approximation factor of $\frac{16\pi}{27}$. Assuming the
Unique Games Conjecture no polynomial time algorithm achieves an
approximation guarantee smaller than $\frac{16\pi}{27}$ in this
case.
\end{theorem}

In fact, we believe that the UGC hardness threshold for the kernel
clustering problem when $A$ is centered and $B$ is spherical and
centered is exactly
$$
\frac{8\pi}{9}\left(1-\frac{1}{k}\right).
$$
In Section~\ref{sec:UGC} we describe a geometric conjecture which we
show implies this tight UGC threshold for general $k$.

We end the discussion by stating a (probably suboptimal) constant
factor approximation result when $A$ is not necessarily centered
(note that this case is of lesser interest in terms of the
applications in machine learning). In this case the above algorithm
gives a constant factor approximation. The slightly better bound on
the approximation factor in Theorem~\ref{thm:noncentered} below
follows from a variant of the above algorithm which will be
described in its proof.

\begin{theorem}\label{thm:noncentered} For general $A$  and $B$ (not necessarily centered)
there exists a polynomial time algorithm that achieves an
approximation factor of
\begin{eqnarray*}
1+\frac{2\pi}{\|v_p-v_q\|_2^2}\cdot \max_{i\in
\{1,\ldots,k\}}\left\|v_i-\frac{v_p+v_q}{2}\right\|_2^2\le
1+\frac{3\pi}{2}.
\end{eqnarray*}
\end{theorem}

The proof of Theorem~\ref{thm:tight UGC} is contained in
Section~\ref{sec:UGC}. We shall  now proceed to prove
Theorem~\ref{thm:analysis}. Before doing so we will show how the
general bound in~\eqref{eq:ugly} implies the
bounds~\eqref{eq:centered} and~\eqref{eq:spherical}. The proof of
Theorem~\ref{thm:noncentered} is deferred to the end of this
section.

To prove that~\eqref{eq:ugly} implies~\eqref{eq:centered} let $D$
denote the diameter of the set $\{v_1,\ldots,v_k\}$, i.e.
$D=\|v_p-v_q\|_2$. A classical theorem of Jung~\cite{Jung1901}
(see~\cite{DGK63}) says that
$$
R(B)\le D\cdot \sqrt{\frac{k-1}{2k}},
$$
and~\eqref{eq:centered} follows immediately by taking the first term
in the minimum in~\eqref{eq:ugly}.

We shall now show that~\eqref{eq:ugly} implies~\eqref{eq:spherical}
when $B$ is either centered and spherical or the identity matrix.
Assume first of all that $B$ is centered and spherical. Note that
since $v_1,\ldots,v_k$ are unit vectors, $R(B)\le 1$. Hence, by
considering the second term in the minimum in~\eqref{eq:ugly} we see
that it is enough to show that there exist $\alpha,\beta,\gamma\in
\{1,\ldots,k\}$ for which
$$
\left\|v_\alpha-\frac{v_\alpha+v_\beta+v_\gamma}{3}\right\|_2^2+\left\|v_\beta-
\frac{v_\alpha+v_\beta+v_\gamma}{3}\right\|_2^2+
\left\|v_\gamma-\frac{v_\alpha+v_\beta+v_\gamma}{3}\right\|_2^2\ge
\frac{2k}{k-1}.
$$
This follows from an averaging argument. Indeed,
\begin{multline*}
\frac{1}{\binom{k}{3}}\sum_{\substack{\alpha,\beta,\gamma\in
\{1,\ldots,k\}\\\alpha<\beta<\gamma}}
\left(\left\|v_\alpha-\frac{v_\alpha+v_\beta+v_\gamma}{3}\right\|_2^2+\left\|v_\beta-
\frac{v_\alpha+v_\beta+v_\gamma}{3}\right\|_2^2+
\left\|v_\gamma-\frac{v_\alpha+v_\beta+v_\gamma}{3}\right\|_2^2\right)\\
=\frac{2}{k}\sum_{i=1}^k
\|v_i\|_2^2-\frac{2}{k(k-1)}\sum_{\substack{i,j\in
\{1,\ldots,k\}\\i\neq j}} \langle
v_i,v_j\rangle=\frac{2}{k}\sum_{i=1}^k
b_{ii}-\frac{2}{k(k-1)}\left(\sum_{i,j=1}^k b_{ij}-\sum_{i=1}^k
b_{ii}\right)=\frac{2k}{k-1}.
\end{multline*}
This complete the proof of~\eqref{eq:spherical} when $B$ is
spherical and centered. The same bound holds true when $B=I_k$ is
the identity matrix since in this case if we denote by $e_1,\ldots,
e_k$ the standard unit basis of $\R^k$ and
$e=\frac{1}{k}\sum_{i=1}^k e_i$ then for every assignment
$\lambda:\{1,\ldots,n\}\to \{1,\ldots,k\}$ we have
\begin{multline}\label{eq:subtract}
\sum_{i,j=1}^n a_{ij}(I_k)_{\lambda(i)\lambda(j)}=\sum_{i,j=1}^n
\langle u_i,u_j\rangle\langle
e_{\lambda(i)},e_{\lambda(j)}\rangle\\=\sum_{i,j=1}^n \langle
u_i,u_j\rangle\langle
e_{\lambda(i)}-e,e_{\lambda(j)}-e\rangle+2\left\langle \sum_{i=1}^n
u_i,\sum_{j=1}^n \langle e,e_{\lambda(j)}\rangle
u_j\right\rangle-\|e\|_2^2 \left\|\sum_{i=1}^k u_i\right\|_2^2.
\end{multline}
The last two terms in~\eqref{eq:subtract} vanish since $A$ is
centered. Thus
$$
\sum_{i,j=1}^n
a_{ij}(I_k)_{\lambda(i)\lambda(j)}=\frac{k-1}{k}\sum_{i,j=1}^n
a_{ij}c_{\lambda(i)\lambda(j)},
$$
where $C=(c_{ij})=\frac{k}{k-1}(\langle e_i-e,e_j-e\rangle)$ is
spherical and centered. Thus the case of the identity matrix reduces
to the previous analysis.

\begin{proof}[Proof of Theorem~\ref{thm:analysis}]
Denote
\begin{eqnarray*}\SDP\coloneqq \max
  \sum_{i,j=1}^n a_{ij} \cdot  \left\langle
   \|w\|_2 u + R(B) x_i, \ \|w\|_2 u + R(B) x_j \right\rangle,
\end{eqnarray*}
where the maximum is taken over all $u,x_1,\ldots,x_n\in \R^{n+1}$
such that $\|u\|_2=1$ and  $\|x_i\|_2\le 1$ for all $i$. Observe
that
\begin{eqnarray}\label{eq:is relaxtion}
\SDP\ge \Clust(A|B).
\end{eqnarray}
Indeed, for every $\lambda:\{1,\ldots,n\}\to \{1,\ldots,k\}$ define
$u=\frac{w}{\|w\|_2}$ and $x_i = \frac{v_{\lambda(i)}-w}{R(B)}$ and
note that in this case
$$
 \sum_{i,j=1}^n a_{ij} \cdot  \left\langle
   \|w\|_2 u + R(B) x_i, \ \|w\|_2 u + R(B) x_j
   \right\rangle=\sum_{i,j=1}^n a_{ij} b_{\lambda(i)\lambda(j)}.
$$

Let $u^*, x_1^*,\ldots, x_n^*$ be the optimal solution to the SDP.
It will be convenient to think of the SDP solution
 as being split into two parts. So we rewrite
 \begin{eqnarray}
 {\rm SDP} & = &   \sum_{i,j=1}^n a_{ij} \cdot  \left\langle
   \|w\|_2 u^* + R(B) x_i^*, \ \|w\|_2 u^* + R(B) x_j^* \right\rangle \nonumber \\ & = &
  \sum_{i,j=1}^n \langle u_i, u_j \rangle \cdot   \langle
   \|w\|_2 u^* + R(B) x^*_i, \ \|w\|_2 u^* + R(B) x_j^* \rangle \nonumber \\
 & = & \left\| \sum_{i=1}^n u_i \otimes (\|w\|_2 u^* + R(B) x_i^*) \right\|_2^2 \\
 & = & \left\| \left( \|w\|_2 \left( \sum_{i=1}^n u_i\right) \otimes u^*
 \right) \label{eq:convex}
  \ + \ \left( R(B)  \sum_{i=1}^n u_i \otimes x_i^* \right) \right\|_2^2 \nonumber \\
  & =& \| P + Q \|_2^2,  \label{eqn:sdp-parts}
\end{eqnarray}
where
\begin{eqnarray}\label{eq:P}
P\coloneqq \|w\|_2 \sum_{i=1}^n u_i\otimes u^*,
\end{eqnarray}
and
\begin{eqnarray}\label{eq:Q}
Q\coloneqq R(B)  \sum_{i=1}^n u_i \otimes x_i^*.
\end{eqnarray}
Observe in passing that~\eqref{eq:convex} implies that the objective
function of the SDP is convex as a function of $u,x_1,\ldots,x_n$,
and therefore we may assume that $\|u^*\|_2=1$ and $\|x_i^*\|_2=1$
for all $i$.

We shall now proceed with the analysis of our algorithm while using
the variant described in Remark~\ref{rem:s}. This will not create
any additional complication, and will allow us to explain why there
is no advantage in working with subsets of size $s\ge 4$. Recall the
setting: for a fixed integer $s\in [2,k]$ we choose a subset
$S\subseteq \{1,..,k\}$ of cardinality $s$ which maximizes the
quantity
$$
\sum_{i\in S} \left\|v_i-\frac{1}{s}\sum_{j\in S} v_j\right\|_2^2.
$$
Then, we choose $s$ i.i.d. standard Gaussians $\{g_i\}_{i\in
S}\subseteq \R^{n+1}$ and define $\sigma:\{1,\ldots,n\}\to
\{1,\ldots,k\}$ by setting $\sigma(r)=i$ if
$$
\langle g_i,x_r^*\rangle=\max_{j\in S} \langle g_j,x_r^*\rangle.
$$
Fix $i,j\in \{1,\ldots, n\}$. As proved by Frieze and Jerrum
in~\cite{FJ97} (see Lemma 5 there), we have\footnote{We are using
here the fact that $x_1^*,\ldots,x_n^*$ are unit vectors.}:
$$
\Pr\left[\sigma(i)=\sigma(j)\right]=\sum_{m=0}^\infty R_m(s) \langle
x_i^*,x_j^*\rangle^m,
$$
where the power series converges on $[-1,1]$ and all the
coefficients $R_m(s)$ are non-negative. Moreover
$R_0(s)=\frac{1}{s}$ and
$$
R_1(s)=\frac{1}{s-1} \left( {\E} \left[  \max_{j \in S} g_j  \right]
\right)^2= \frac{s}{(2\pi)^{s/2}}\int_{-\infty}^\infty
xe^{-x^2/2}\left(\int_{-\infty}^x e^{-y^2/2}dy\right)^{s-1}dx.
$$

Note that conditioned on the event $\sigma(i)=\sigma(j)$, the random
index $\sigma(i)$ is uniformly distributed over $S$. Also,
conditioned on the event $\sigma(i)\neq \sigma(j)$, the pair
$(\sigma(i),\sigma(j))$ is uniformly distributed over all $s(s-1)$
pairs of distinct indices in $S$. Thus
 $$ \E [ b_{\sigma(i)\sigma(j)} ] =
  {\rm Pr}[\sigma(i) = \sigma(j)] \cdot
  \left( \frac{1}{s}\sum_{\ell\in S} b_{\ell \ell} \right) +
    {\rm Pr}[\sigma(i) \not= \sigma(j)] \cdot
  \left( \frac{1}{s(s-1)}\sum_{\substack{\ell,t\in S\\\ell\neq t}}
   b_{\ell t} \right). $$

   Denote $\Phi = \frac{1}{s}\sum_{\ell\in S} b_{\ell \ell}$ and $\Psi =
\frac{1}{s(s-1)}\sum_{\substack{\ell,t\in S\\\ell\neq t}}
   b_{\ell t}$. (note that $\Phi, \Psi$ depend on the matrix $B$ as well as the
choice of the subset $S\subseteq \{1,\ldots,k\}$.     Thus
\begin{multline}
  \E [ b_{\sigma(i)\sigma(j)} ]  =
  \left( \sum_{m=0}^\infty R_m(s) \langle
x_i^*,x_j^*\rangle^m \right) \cdot
  \Phi
   \ + \ \left( 1- \sum_{m=0}^\infty R_m(s) \langle
x_i^*,x_j^*\rangle^m \right) \cdot
   \Psi \\
    = \left(\Psi + (\Phi - \Psi) R_0(s)\right) + (\Phi - \Psi )
    \sum_{m=1}^\infty R_m(s) \langle
x_i^*,x_j^*\rangle^m.  \label{eqn:rounding-1}
\end{multline}

Write $v \coloneqq \frac{1}{s} \sum_{\ell\in S} v_\ell$. Observe
that
\begin{eqnarray}\label{eq:const-term}
\Psi+(\Phi-\Psi)R_0(s)=\|v\|_2^2. \end{eqnarray} Indeed, since
$R_0(s)=1/s$ we have
$$
\Psi+(\Phi-\Psi)R_0(s)=\left(1-\frac{1}{s}\right) \left(
\frac{1}{s(s-1)}\sum_{\substack{\ell,t\in S\\\ell\neq t}}
   b_{\ell t}  \right) +
 \frac{1}{s} \left(\frac{1}{s}\sum_{\ell\in S} b_{\ell \ell}\right) =
 \frac{1}{s^2}
  \sum_{\ell,t\in S} b_{\ell t} = \left\| \frac{1}{s} \sum_{\ell\in S}^s v_\ell \right\|_2^2
  = \| v \|_2^2.
$$
Moreover,
\begin{eqnarray}\label{eq:alpha-beta}
(s-1)(\Phi - \Psi) = \sum_{\ell\in S} \| v_\ell - v \|_2^2.
\end{eqnarray}
In particular $\Phi-\Psi\ge 0$. To prove~\eqref{eq:alpha-beta} we
simply expand:
$$
\sum_{\ell\in S} \| v_\ell - v \|_2^2=\sum_{\ell\in S}
\|v_\ell\|_2^2-s\|v\|_2^2=s\Phi-\frac{1}{s}\sum_{\ell,t\in S}b_{\ell
t}=s\Phi-\frac{1}{s}\left(s\Phi+s(s-1)\Psi\right)=(s-1)(\Phi -
\Psi).
$$

Multiplying both sides of equation~\eqref{eqn:rounding-1} by
$a_{ij}$ and summing over $i,j \in \{1,\ldots,n\}$ while
using~\eqref{eq:const-term} we get that
\begin{eqnarray}\label{eq:rounded}\E \left[ \sum_{i,j=1}^n a_{ij}
b_{\sigma(i) \sigma(j)}\right]
   =  \|v\|_2^2 \sum_{i,j=1}^n a_{ij}  +
   (\Phi - \Psi) R_1(s) \sum_{i,j=1}^n a_{ij} \langle x_i^*,x_j^*\rangle  +
  (\Phi- \Psi) \sum_{m = 2}^\infty R_m(s)  \sum_{i,j=1}^n
   a_{ij} \langle x_i^*,x_j^*\rangle^m .  \end{eqnarray}
Note that for every $m\ge 1$ we have
\begin{eqnarray}\label{eq:higher terms}
\sum_{i,j=1}^n
   a_{ij} \langle x_i^*,x_j^*\rangle^m =\sum_{i,j=1}^n \langle
   u_i,u_j\rangle \left\langle (x_i^*)^{\otimes m}, (x_j^*)^{\otimes
   m}\right\rangle= \left\|\sum_{i=1}^nu_i\otimes (x_i^*)^{\otimes
   m}\right\|_2^2\ge 0.
\end{eqnarray}
Plugging~\eqref{eq:higher terms} into~\eqref{eq:rounded}, and using
the fact that $\Phi-\Psi\ge 0$ and the positivity of $\R_m(s)$, we
conclude that
\begin{eqnarray}\label{eq:conclusion}
\E \left[ \sum_{i,j=1}^n a_{ij} b_{\sigma(i) \sigma(j)}\right]\ge
\|v\|_2^2 \sum_{i,j=1}^n a_{ij}  +
   (\Phi - \Psi) R_1(s) \sum_{i,j=1}^n a_{ij} \langle
   x_i^*,x_j^*\rangle.
\end{eqnarray}

We shall now use the fact that $\sum_{i,j=1}^na_{ij}=0$ for the
first time. In this case $P=0$ (see equations~\eqref{eqn:sdp-parts}
and~\eqref{eq:P}) so that
\begin{eqnarray}\label{eq:now centered}
\SDP= R(B)^2\sum_{i,j=1}^na_{ij}\langle x_i^*,x_j^*\rangle.
\end{eqnarray}
Hence, using~\eqref{eq:conclusion} and~\eqref{eq:const-term} we get
the bound
\begin{eqnarray}\label{eq:almost done}
 \E \left[ \sum_{i,j=1}^n a_{ij} b_{\sigma(i)
\sigma(j)}\right]\ge \frac{R_1(s)\sum_{\ell\in S} \| v_\ell - v
\|_2^2}{(s-1)R(B)^2}\cdot \SDP\stackrel{\eqref{eq:is
relaxtion}}{\ge}\frac{R_1(s)\sum_{\ell\in S} \| v_\ell - v
\|_2^2}{(s-1)R(B)^2}\cdot \Clust(A|B).
\end{eqnarray}
The term $R_1(s)$ is studied in Section~\ref{sec:geometry}, where
its geometric interpretation is explained. In particular, it follows
from Corollary~\ref{coro:not simplex} and Corollary~\ref{coro:2,3}
that $R_1(s)< R_1(3)$ for every $s\ge 4$ and that
$R_1(2)=\frac{1}{\pi}$ and $R_1(3)=\frac{9}{8\pi}$. Hence the cases
$s\in \{2,3\}$ in~\eqref{eq:almost done} conclude the proof of
Theorem~\ref{thm:analysis}. Moreover, we see that for $s\ge 4$ the
lower bound in~\eqref{eq:almost done} is worse than the lower bound
obtained when case $s=3$. Indeed, we have already noted that in this
case $R_1(s)<R_1(3)$. In addition,
$$
\frac{1}{\binom{s}{3}} \sum_{\substack{T\subseteq S\\ |T|=3}}
\frac{1}{2} \sum_{\ell\in T} \left\|v_\ell-\frac{1}{3}\sum_{t\in T}
v_t\right\|_2^2=\frac{1}{s}\sum_{\ell\in S}
\|v_\ell\|_2^2-\frac{1}{s(s-1)}\sum_{\substack{\ell,t\in S\\\ell\neq
t}} \langle v_\ell,v_t\rangle=\frac{1}{s-1} \sum_{\ell\in S}
\left\|v_\ell-\frac{1}{s}\sum_{t\in S} v_t\right\|_2^2.
$$
This implies that there exists $T\subseteq S$ with $|T|=3$ for which
$$
\frac{1}{2} \sum_{\ell\in T} \left\|v_\ell-\frac{1}{3}\sum_{t\in T}
v_t\right\|_2^2\ge \frac{1}{s-1}\sum_{\ell\in S} \| v_\ell - v
\|_2^2,
$$
so that when $s\ge 4$ the lower bound in~\eqref{eq:almost done} is
inferior to the same lower bound when $s=3$.
\end{proof}

It remains to deal with the case $\sum_{i,j=1}^n a_{ij}>0$, i.e. to
prove Theorem~\ref{thm:noncentered}.

\begin{proof}[Proof of Theorem~\ref{thm:noncentered}] We slightly
modify the algorithm that was studied in Theorem~\ref{thm:analysis}.
Let $v_1,\ldots, v_k$ and $p,q\in \{1,\ldots,k\}$ be as before, that
is $b_{ij}=\langle v_i,v_j\rangle $ and $\|v_p-v_q\|_2=\max_{i,j\in
\{1,\ldots,k\}}\|v_i-v_j\|_2=D$, the diameter of the set
$\{v_1,\ldots,v_k\}\in \R^k$. Denote $w'\coloneqq\frac{v_p+v_q}{2}$
and
$$
R'(B)\coloneqq \max_{i\in \{1,\ldots,k\}}\left\|v_i-w'\right\|_2.
$$
We now consider the modified semidefinite program
\begin{eqnarray*}\SDP\coloneqq \max
  \sum_{i,j=1}^n a_{ij} \cdot  \left\langle
   \|w'\|_2 u + R'(B) x_i, \ \|w'\|_2 u + R'(B) x_j \right\rangle,
\end{eqnarray*}
where the maximum is taken over all $u,x_1,\ldots,x_n\in \R^{n+1}$
such that $\|u\|_2=1$ and  $\|x_i\|_2\le 1$ for all $i$. From now on
we will use the notation of the proof of Theorem~\ref{thm:analysis}
with $w$ replaced by $w'$ and $R(B)$ replaced by $R'(B)$ (this
slight abuse of notation will not create any confusion). As before,
we let $g_1,g_2\in \R^{n+1}$ be i.i.d. standard Gaussian vectors and
define $\sigma: \{1,\ldots,n\}\to \{1,\ldots,k\}$ by
\begin{eqnarray}\label{eq:def sigma}
\sigma(r)=\left\{\begin{array}{ll} p & \mathrm{if}\ \langle
g_1,x_r\rangle \ge \langle g_2,x_r\rangle,\\
q & \mathrm{if}\ \langle g_2,x_r\rangle \ge \langle g_1,x_r\rangle.
\end{array}\right.
\end{eqnarray}
Note that the first place in the proof of Theorem~\ref{thm:analysis}
where the assumption that $A$ is centered was used in
equation~\eqref{eq:now centered}. Hence, in the present setting we
still have the bounds
\begin{eqnarray}\label{eq:new clust bound}
\Clust(A|B)\le \SDP=\|P+Q\|_2^2\le \left(\|P\|_2+\|Q\|_2\right)^2,
\end{eqnarray}
where $P$ and $Q$ are defined in~\eqref{eq:P} and~\eqref{eq:Q} (with
$w$ and $R(B)$ replaced by $w'$ and $R'(B)$, respectively). Also, it
follows from~\eqref{eq:conclusion} that
\begin{eqnarray}\label{eq:conclusion2}
\E \left[ \sum_{i,j=1}^n a_{ij} b_{\sigma(i) \sigma(j)}\right]\ge
\|v\|_2^2 \sum_{i,j=1}^n a_{ij}  +
   \left(\|v_p-v\|_2^2+\|v_q-v\|_2^2\right) R_1(2) \sum_{i,j=1}^n a_{ij} \langle
   x_i^*,x_j^*\rangle,
\end{eqnarray}
where $v=\frac{v_p+v_q}{2}=w'$. Note that
$\|v_p-v\|_2^2+\|v_q-v\|_2^2=\frac{D^2}{2}$, and recall that
$R_1(2)=\frac{1}{\pi}$. Thus~\eqref{eq:conclusion2} becomes:
\begin{eqnarray}\label{eq:conclusion3}
\E \left[ \sum_{i,j=1}^n a_{ij} b_{\sigma(i) \sigma(j)}\right]\ge
\|w'\|_2^2 \sum_{i,j=1}^n a_{ij}  +
 \frac{D^2}{2\pi} \sum_{i,j=1}^n a_{ij} \langle
   x_i^*,x_j^*\rangle,
\end{eqnarray}
Note that
\begin{eqnarray}\label{eq:Pnew}
\|P\|_2^2= \|w'\|_2^2\cdot\left\|\sum_{i=1}^n u_i\otimes
u^*\right\|_2^2=\|w'\|_2^2\cdot \|u^*\|_2^2\sum_{i,j=1}^n\langle
u_i,u_j\rangle=\|w'\|_2^2\sum_{i,j=1}^n a_{ij}.
\end{eqnarray}
and
\begin{eqnarray}\label{eq:Qnew}
\|Q\|_2^2=R'(B)^2\cdot\left\|\sum_{i=1}^n u_i\otimes
x_i^*\right\|_2^2=R'(B)^2\sum_{i,j=1}^n a_{ij}\langle
x_i^*,x_j^*\rangle.
\end{eqnarray}
Combining~\eqref{eq:new clust bound} and~\eqref{eq:conclusion3}
with~\eqref{eq:Pnew} and~\eqref{eq:Qnew} we see that
\begin{eqnarray}\label{eq:end noncentered}
\Clust(A|B)\le
\frac{\left(\|P\|_2+\|Q\|_2\right)^2}{\|P_2\|_2^2+c\|Q\|_2^2}\cdot
\E \left[ \sum_{i,j=1}^n a_{ij} b_{\sigma(i) \sigma(j)}\right],
\end{eqnarray}
where $c=\frac{D^2}{2\pi R'(B)^2}$. The convexity of the function
$x\to x^2$ implies that
\begin{multline*}
\left(\|P\|_2+\|Q\|_2\right)^2=\left(\frac{c}{c+1}\cdot
\frac{c+1}{c}\|P\|_2^2+\left(1-\frac{c}{c+1}\right)(c+1)\|Q\|_2^2\right)^2\\\le
\frac{c+1}{c}\|P\|_2^2+(c+1)\|Q\|_2^2=\left(1+\frac{1}{c}\right)\left(\|P\|_2^2+c\|Q\|_2^2\right).
\end{multline*}
Thus~\eqref{eq:end noncentered} implies that our algorithm achieves
an approximation guarantee bounded above by
$$
1+\frac{1}{c}=1+ \frac{2\pi R'(B)^2}{D^2}=
1+\frac{2\pi}{\|v_p-v_q\|_2^2}\cdot \max_{i\in
\{1,\ldots,k\}}\left\|v_i-\frac{v_p+v_q}{2}\right\|_2^2.
$$
It remains to note that for every $i\in \{1,\ldots,k\}$ we know that
$\|v_i-v_p\|_2,\|v_i-v_q\|_2\le D$ and therefore $\|v_i-w'\|_2\le
\frac{\sqrt{3}}{2}D$. This implies that our approximation guarantee
is bounded from above by $1+\frac{3\pi}{2}$.
\end{proof}

\section{UGC hardness}\label{sec:UGC}

\subsection{Geometric preliminaries: Propeller problems}\label{sec:geometry}

Let $\gamma_n$ be the standard Gaussian measure on $\R^n$. For any
integer $k\ge 2$ define
\begin{eqnarray}\label{eq:defC}
C(n,k)\coloneqq \sup\left\{\sum_{j=1}^k
\left\|\int_{\R^n}xf_j(x)d\gamma_n(x)\right\|_2^2:\
f_1,\ldots,f_k\in L_2(\gamma_n) \ \wedge \ \forall j\  f_j\ge 0\
\wedge  \ \sum_{j=1}^k f_j\le 1\right\}.
\end{eqnarray}
We first observe that the supremum in~\eqref{eq:defC} is attained at
a $k$-tuple of functions which correspond to a partition of $\R^n$:
 \begin{lemma}\label{lem:maximizer exists} There exist disjoint measurable sets $A_1,\ldots,A_k\subseteq
 \R^n$ such that $A_1\cup A_2\cup\cdots \cup A_k=\R^n$ and
$$
\sum_{j=1}^k \left\|\int_{A_j}xd\gamma_n(x)\right\|_2^2=C(n,k).
$$
 \end{lemma}

\begin{proof}
Let $H$ be the Hilbert space $L_2(\gamma_n)\oplus
L_2(\gamma_n)\oplus \cdots \oplus L_2(\gamma_n)$ ($k$ times). Define
$K\subseteq H$ to be the set of all $(f_1,\ldots,f_k)\in H$ such
that $f_j\ge 0$ for all $j$ and $\sum_{j=1}^k f_j\le 1$. Then $K$ is
a closed convex and bounded subset of $H$, and hence by the
Banach-Alaoglu it is weakly compact. The mapping $\psi:K\to \R$
given by
$$
\psi(f_1,\ldots,f_k)\coloneqq \sum_{j=1}^k
\left\|\int_{\R^n}xf_j(x)d\gamma_n(x)\right\|_2^2
=\sum_{j=1}^k\sum_{i=1}^n \left(\int_{\R^n} x_i
f_j(x)d\gamma_n(x)\right)^2
$$
is weakly continuous since the mapping $(x_1,\ldots,x_n)\to x_j$ is
in $L_2(\gamma_n)$ for each $j$. Hence $\psi$ attains its maximum on
$K$, say at $(f_1^*,\ldots,f_k^*)\in K$.

Define $z_j\coloneqq \int_{\R^n} x f_j^*(x)d\gamma_n(x)\in \R^n$ and
let
$$
w\coloneqq -\sum_{j=1}^kz_j= \int_{\R^n} x
\left(1-\sum_{j=1}^kf_j^*(x)\right)d\gamma_n(x).
$$
Note that
$$
\frac{1}{k} \sum_{i=1}^k\left(\sum_{\substack{1\le j\le k\\j\neq
i}}\|z_j\|_2^2+\|z_i+w\|_2^2\right)= \sum_{j=1}^k \|z_j\|_2^2
+\left(1-\frac{2}{k}\right)\|w\|_2^2\ge \sum_{j=1}^k \|z_j\|_2^2,
$$
which implies the existence of  $i\in \{1,\ldots, k\}$ for which
$$
\sum_{\substack{1\le j\le k\\j\neq i}}\|z_j\|_2^2+\|z_i+w\|_2^2\ge
\sum_{j=1}^k \|z_j\|_2^2.
$$
Hence, if we define for $j\in \{1,\ldots, k\}$,
$$
g_j\coloneqq \left\{\begin{array}{ll} f_j^* & j\neq i\\
f_i^*+ 1-\sum_{j=1}^kf_j^*& j=i\end{array}\right.
$$
then $(g_1,\ldots,g_k)\in K$, and
$$
C(n,k)\ge \sum_{j=1}^k
\left\|\int_{\R^n}xg_j(x)d\gamma_n(x)\right\|_2^2=\sum_{\substack{1\le
j\le k\\j\neq i}}\|z_j\|_2^2+\|z_i+w\|_2^2\ge \sum_{j=1}^n
\|z_j\|_2^2=C(n,k).
$$
So
$$
\sum_{j=1}^k\left\|\int_{\R^n}xg_j(x)d\gamma_n(x)\right\|_2^2=C(n,k).
$$
Note that $\sum_{j=1}^kg_j=1$, so we can define a random partition
$A_1,\ldots,A_k$ of $\R^n$ as follows: let $\{s_x\}_{x\in \R^n}$ be
independent random variables taking values in $\{1,\ldots,k\}$ such
that $\Pr(s_x=j)=g_j(x)$, and define $A_j\coloneqq \{x\in \R^n:\
s_x=j\}$. Then by convexity and the definition of $C(n,k)$ we see
that
$$
\E\sum_{j=1}^k \left\|\int_{A_j}xd\gamma_n(x)\right\|_2^2\ge
\sum_{j=1}^k
\left\|\int_{\R^n}\big(\E\1_{A_j}(x)\big)xd\gamma_n(x)\right\|_2^2=
\sum_{j=1}^k\left\|\int_{\R^n}xg_j(x)d\gamma_n(x)\right\|_2^2=C(n,k).
$$
It therefore follows that there exists a partition as required.
\end{proof}

\begin{lemma}\label{lem:C(k-1,k)}
If $n\ge k-1$ then $C(n,k)=C(k-1,k)$ and if $n<k-1$ then
$C(n,k)=C(n,n+1)$.
\end{lemma}

\begin{proof}
Assume first of all that $n\ge k-1$. The inequality $C(n,k)\ge
C(k-1,k)$ is easy since for every $f_1,\ldots,f_k\in
L_2(\gamma_{k-1})$ which satisfy $f_j\ge 0$ for all $j\in
\{1,\ldots,k\}$ and $f_1+\cdots+f_k\le 1$ we can define $\widetilde
f_1,\ldots\widetilde f_k:\R^n=\R^{k-1}\times \R^{n-k+1}\to \R$ by
$\widetilde f_j(x,y)=f_j(x)$. Then $\widetilde f_1,\ldots\widetilde
f_k\in L_2(\gamma_n)$, $\widetilde f_1,\ldots\widetilde f_k\ge 0$,
$\widetilde f_1+\cdots+\widetilde f_k\le 1$ and $\sum_{j=1}^k
\left\|\int_{\R^{k-1}}xf_j(x)d\gamma_{k-1}(x)\right\|_2^2=\sum_{j=1}^k
\left\|\int_{\R^n}x\widetilde f_j(x)d\gamma_n(x)\right\|_2^2$. In
the reverse direction, by Lemma~\ref{lem:maximizer exists} there is
a measurable partition $A_1,\ldots,A_k$ of $\R^n$ such that if we
define $z_j\coloneqq \int_{A_j}xd\gamma_n(x)\in \R^n$ then we have
$\sum_{j=1}^k \left\|z_j\right\|_2^2=C(n,k)$. Note that
$$
\sum_{j=1}^k z_j=\int_{\R^n}\left(\sum_{j=1}^k \1_{A_j}\right)
xd\gamma_n(x)=\int_{\R^n} xd\gamma_n(x)=0.
$$
Hence the dimension of the subspace $V\coloneqq
\mathrm{span}\{z_1,\ldots,z_k\}$ is $d\le k-1$. Define
$g_1,\ldots,g_k:V\to [0,1]$ by
$$
g_j(x)=\gamma_{V^{\perp}}\left((A_j-x)\cap V^\perp\right).
$$
Then $g_1+\cdots+g_k=1$, so that
\begin{multline*}
C(k-1,k)\ge C(d,k)\ge \sum_{j=1}^k \left\|\int_{V}
xg_j(x)d\gamma_V(x)\right\|_2^2=\sum_{j=1}^k
\left\|\int_{V}\int_{V^\perp}\1_{A_j}(x+y)
xd\gamma_V(x)d\gamma_{V^\perp}(y)\right\|_2^2\\
= \sum_{j=1}^k
\left\|\int_{A_j}\mathrm{Proj}_{V}(w)d\gamma_n(w)\right\|_2^2=\sum_{j=1}^k
\left\|\mathrm{Proj}_{V}(z_j)\right\|_2^2=\sum_{j=1}^k
\left\|z_j\right\|_2^2=C(n,k).
\end{multline*}

We now pass to the case $n<k-1$. The inequality $C(n,n+1)\le C(n,k)$
is trivial, so we need to show that $C(n,k)\le C(n,n+1)$. We observe
that since $k>n+1$ for every $v_1,\ldots,v_k\in \R^n$ there exist
two distinct indices $i,j\in \{1,\ldots,k\}$ such that $\langle
v_i,v_j\rangle \ge 0$. The proof of this fact is by induction on
$n$. If $n=1$ then our assumption is that $k\ge 3$, and therefore at
least two of the real numbers $v_1,\ldots,v_k$ must have the same
sign. For $n>1$ we may assume that $\langle v_1, v_j\rangle<0$ for
all $j\ge 2$ (otherwise we are done). Consider the vectors
$\left\{v_j-\frac{\langle v_1,v_j\rangle}{\|v_1\|_2^2}\cdot
v_1\right\}_{j=2}^k$, i.e. the projections of $v_2,\ldots,v_k$ onto
the orthogonal complement of $v_1$. By induction there are distinct
$i,j\in \{2,\ldots,k\}$ such that
$$
0\le \left\langle v_i-\frac{\langle
v_1,v_i\rangle}{\|v_1\|_2^2}\cdot v_1, v_j-\frac{\langle
v_1,v_j\rangle}{\|v_1\|_2^2}\cdot v_1 \right\rangle=\langle
v_i,v_j\rangle -\frac{\langle v_i,v_1\rangle\langle
v_j,v_1\rangle}{\|v_1\|_2^2}\le \langle v_i,v_j\rangle.
$$
Now, let $A_1,\ldots,A_k$ be a partition of $\R^n$ as in
Lemma~\ref{lem:maximizer exists} and denote $z_j\coloneqq
\int_{A_j}xd\gamma_n(x)\in \R^n$. By the above argument there are
distinct $i,j\in \{1,\ldots,k\}$ such that $\langle
z_i,z_j\rangle\ge 0$. Hence
\begin{multline*}
C(n,k-1)\ge \sum_{\substack{1\le \ell\le k\\\ell
\notin\{i,j\}}}\left\|\int_{A_\ell}xd\gamma_n(x)\right\|_2^2+\left\|\int_{A_i\cup
A_j}xd\gamma_n(x)\right\|_2^2=\sum_{\substack{1\le \ell\le k\\\ell
\notin\{i,j\}}} \|z_\ell\|_2^2+\|z_i+z_j\|_2^2\\\ge
\sum_{\ell=1}^k\|z_\ell\|_2^2=C(n,k)\ge C(n,k-1).
\end{multline*}
So, $C(n,k)=C(n,k-1)$, and the required identity follows by
induction.
\end{proof}

In light of Lemma~\ref{lem:C(k-1,k)} we denote from now on
$C(k)\coloneqq C(k-1,k)$. Given distinct $z_1,\ldots,z_k\in
\R^{k-1}$ and $j\in \{1,\ldots, k\}$ define a set
$P_j(z_1,\ldots,z_k)\subseteq \R^k$ by
$$
P_j(z_1,\ldots,z_k)\coloneqq \left\{x\in \R^{k}:\ \langle
x,z_j\rangle =\max_{i\in \{1,\ldots,k\}} \langle
x,z_i\rangle\right\}.
$$
Thus $\left\{P_j(z_1,\ldots,z_k)\right\}_{j=1}^k$ is a partition of
$\R^{k-1}$  which we call the simplicial partition induced by
$z_1,\ldots,z_k$ (strictly speaking the elements of this partition
are not disjoint, but they intersect at sets of measure $0$).

\begin{lemma}\label{lem:simplicial}
Let $A_1,\ldots,A_k\subseteq \R^{k-1}$ be a partition as in
Lemma~\ref{lem:maximizer exists}, i.e. if we set $z_j\coloneqq
\int_{A_j}xd\gamma_{k-1}(x)$ then $C(k)=\sum_{j=1}^k \|z_j\|_2^2$.
Assume also that this partition is minimal in the sense that the
number of elements of positive measure in this partition is minimal
among all the possible partitions from Lemma~\ref{lem:maximizer
exists}. By relabeling we may assume without loss of generality that
for some $1\le \ell\le k$ we have
$\gamma_{k-1}(A_1),\ldots,\gamma_{k-1}(A_\ell)>0$ and that
$\gamma_{k-1}(A_{\ell+1})=\cdots=\gamma_{k-1}(A_k)=0$. Then up to an
orthogonal transformation $z_1,\ldots,z_\ell\in \R^{\ell-1}$, for
any distinct $i,j\in \{1,\ldots,\ell\}$ we have $\langle
z_i,z_j\rangle<0$ and for each $j\in\{1,\ldots, \ell\}$ we have
$A_j=P_j(z_1,\ldots,z_\ell)\times \R^{k-\ell}$ up to sets of measure
zero.
\end{lemma}

\begin{proof} Since $\1_{A_1}+\cdots+\1_{A_\ell}=1$ almost
everywhere we have $z_1+\cdots z_\ell=0$. Thus the dimension of the
span of $z_1,\ldots,z_\ell$ is at most $\ell-1$, and by applying an
orthogonal transformation we may assume that $z_1,\ldots,z_\ell\in
\R^{\ell-1}$. Also, if for some distinct $i,j\in \{1,\ldots,\ell\}$
we have $\langle z_i,z_j\rangle \ge 0$ we may replace $A_i$ by
$A_i\cup A_j$ and $A_j$ by the empty set and obtain a partition of
$\R^{k-1}$ which contains exactly $\ell-1$ elements of positive
measure and for which we have:
\begin{eqnarray*}
C(k)\ge \sum_{\substack{1\le r\le k\\\ell
\notin\{i,j\}}}\left\|\int_{A_r}xd\gamma_n(x)\right\|_2^2+\left\|\int_{A_i\cup
A_j}xd\gamma_n(x)\right\|_2^2=\sum_{\substack{1\le r\le k\\\ell
\notin\{i,j\}}} \|z_r\|_2^2+\|z_i+z_j\|_2^2\ge
\sum_{r=1}^k\|z_r\|_2^2=C(k).
\end{eqnarray*}
This contradicts the minimality of the partition $A_1,\ldots,A_k$.

Note that the above reasoning implies in particular that the vectors
$z_1,\ldots,z_\ell$ are distinct, and therefore
$\left\{P_j(z_1,\ldots,z_\ell)\times \R^{k-\ell}\right\}_{j=1}^\ell$
is a partition of $\R^{k-1}$ (up to sets of measure $0$). Assume for
the sake of contradiction that these exist
 $i\in \{1,\ldots,\ell\}$ such that
$$
\gamma_{k-1}\left(A_i\setminus \left(P_j(z_1,\ldots,z_\ell)\times
\R^{k-\ell}\right)\right)>0.
$$
Note that up to sets of measure $0$ we
 have:
$$
A_i\setminus \left(P_j(z_1,\ldots,z_\ell)\times
\R^{k-\ell}\right)=\bigcup_{\substack{j\in \{1,\ldots,\ell\}\\j\neq
i}}\bigcup_{m=1}^\infty \left\{x\in A_i:\ \langle x,z_j\rangle \ge
\langle x,z_i\rangle+ \frac{1}{m}\right\}.
$$
Hence there exists $m>0$ and $j\in \{1,\ldots,\ell\}\setminus \{i\}$
such that if we denote $E\coloneqq \left\{x\in A_i:\ \langle
x,z_j\rangle \ge \langle x,z_i\rangle+ \frac{1}{m}\right\}$ then
$\gamma_{k-1}(E)>0$.
Define a
partition $\widetilde A_1,\ldots \widetilde A_k$ of $\R^{k-1}$ by
$$
\widetilde A_r\coloneqq \left\{\begin{array}{ll}A_r &
r\notin \{i,j\}\\
A_i\setminus E& r=i\\
A_j\cup E & r=j.\end{array}\right.
$$
Then for $w\coloneqq \int_{E}xd\gamma_{k-1}(x)$ we have
\begin{multline*}
C(k)\ge \sum_{r=1}^k \left\|\int_{\widetilde A_r}
xd\gamma_{k-1}(x)\right\|_2^2=\sum_{\substack{1\le r\le k\\r
\notin\{i,j\}}} \|z_r\|_2^2+\|z_i-w\|_2^2+\|z_j+w\|_2^2=\sum_{r=1}^k
\|z_r\|_2^2+2\|w\|_2^2+2\langle z_j,w\rangle-2\langle
z_i,w\rangle\\\ge C(k)+2\|w\|_2^2+2\int_{E} \left(\langle
z_j,x\rangle-\langle z_i,x\rangle\right)d\gamma_{k-1}(x)\ge
C(k)+\frac{2\gamma_{k-1}(E)}{m}>C(k),
\end{multline*}
a contradiction.
\end{proof}

\begin{corollary}\label{coro:2,3}
We have $C(2)=\frac{1}{\pi}$ and $C(3)=\frac{9}{8\pi}$.
\end{corollary}

\begin{proof}
Note that Lemma~\ref{lem:simplicial} implies that for each $k\ge 2$
there exists a partition $A_1,\ldots,A_{k}$ of $\R^{k-1}$ such that
each $A_j$ is a cone and $C(k)=\sum_{j=1}^k
\left\|\int_{A_j}xd\gamma_{k-1}(x)\right\|_2^2$. When $k=2$ the only
such partition of $\R$ consists of the positive and negative
half-lines. Thus
$$
C(2)=2\left(\frac{1}{\sqrt{2\pi}}\int_0^\infty
xe^{-x^2/2}dx\right)^2=\frac{1}{\pi}.
$$

When $k=3$ the partition $A_1,A_2,A_3$ consists of disjoint cones of
angles $\alpha_1,\alpha_2,\alpha_3\in [0,2\pi]$, respectively, where
$\alpha_1+\alpha_2+\alpha_3=2\pi$. Now, for $j\in \{1,2,3\}$ we have
$$
\left\|\int_{A_j}xd\gamma_2(x)\right\|_2^2=\left|\frac{1}{2\pi}\int_0^\infty
\int_{-\alpha_j/2}^{\alpha_j/2}e^{i\theta}r^2e^{-r^2/2}dr
d\theta\right|^2=\frac{\sin^2(\alpha_j/2)}{2\pi}.
$$
Hence
\begin{multline}\label{eq:lagrange}
C(3)=\frac{1}{2\pi}\max\left\{\sin^2(\alpha_1/2)+\sin^2(\alpha_2/2)+\sin^2(\alpha_3/2):\
\alpha_1,\alpha_2,\alpha_3\in [0,\pi]\ \wedge\
\alpha_1+\alpha_2+\alpha_3=2\pi\right\}\\
= \frac{3}{2\pi}\cdot
\sin^2\left(\frac{\pi}{3}\right)=\frac{9}{8\pi},
\end{multline}
where~\eqref{eq:lagrange} follows from a simple Lagrange multiplier
argument.
\end{proof}

It is tempting to believe that for every $k\ge 2$, $C(k)$ is
attained at a regular simplicial partition, i.e. a partition of
$\R^{k-1}$ of the form $\{P_j(v_1,\ldots,v_k)\}_{j=1}^k$ where
$v_1,\ldots,v_k$ are the vertices of the regular simplex in
$\R^{k-1}$. This was shown to be true for $k\in \{2,3\}$ in
Corollary~\ref{coro:2,3}. We will now show that this is not the case
for $k\ge 4$.

\begin{lemma}\label{lem:regular identity}
Let $v_1, v_2, \ldots, v_k\in \R^{k-1}$ be vertices of a regular
simplex in $\R^{k-1}$, i.e. for each $i\in \{1,\ldots,k\}$ we have
$\| v_i \|_2=1$ and for each distinct $i,j\in \{1,\ldots,k\}$ we
have $ \langle v_i,v_j \rangle= -\frac{1}{k-1}$. Let
 $$ z_i \coloneqq  \int_{P_i(v_1,\ldots,v_k)} x \ d \gamma_{k-1}(x). $$
Then $$\sum_{i=1}^k \| z_i \|_2^2 =  \frac{1}{k-1} \left( {\E}
\left[  \max_{j \in \{1,\ldots,k\}} g_j  \right] \right)^2,  $$
where $g_1, g_2, \ldots, g_k$ are independent standard Gaussian
random variables.
\end{lemma}

\begin{proof}
By symmetry all the $z_i$ have the same length $r>0$ and $z_i$ has
the same direction as $v_i$. Thus for all $i$ we have $\langle
z_i,v_i\rangle =r$. Now,
\begin{multline*}
\sum_{i=1}^k \langle z_i,v_i\rangle = \sum_{i=1}^k
\int_{P_i(v_1,\ldots,v_k)} \langle x, v_i\rangle  \
d\gamma_{k-1}(x)=\sum_{i=1}^k  \int_{P_i(v_1,\ldots,v_k)} \left(
\max_{j\in \{1,\ldots,k\}}
   \langle x, v_j \rangle\right)  d\gamma_{k-1}(x) \\= \int_{\R^{k-1}}  \left(
\max_{j\in \{1,\ldots,k\}}
   \langle x, v_j \rangle\right)  d\gamma_{k-1}(x)=\E \left[\max_{j\in \{1,\ldots,k\}}
   h_j\right],
\end{multline*}
where $h_1,\ldots, h_k$ are standard Gaussian random variables with
covariances $\E[h_ih_j]=\langle v_i,v_j\rangle$. Let $h$ be a
standard Gaussian which is independent of $h_1,\ldots,h_k$. Then
\begin{eqnarray}\label{eq:add h} {\E} \left[ \max_{j \in \{1,\ldots,k\}} h_j \right] =
    {\E} \left[ \frac{h}{\sqrt{k-1}} + \left(\max_{j \in \{1,\ldots,k\}} h_j\right) \right]
 = {\E} \left[ \max_{j \in \{1,\ldots,k\}} \left( \frac{h}{\sqrt{k-1}} + h_j\right) \right]
 = {\E} \left[ \max_{j \in \{1,\ldots,k\}} \widetilde{h}_j
 \right],\end{eqnarray}
where we set $\widetilde{h}_j \coloneqq \frac{h}{\sqrt{k-1}} + h_j$
so that $\widetilde{h}_j$ are independent Gaussians with mean zero
and variance $\frac{k}{k-1}$. The last term in~\eqref{eq:add h} is
same as
 $\sqrt{\frac{k}{k-1}}\cdot {\E} \left[ \max_{j \in \{1,\ldots,k\}} g_j \right]$ where $g_1,\ldots,g_k$ are
  independent standard Gaussians.
\end{proof}

\begin{corollary}\label{coro:not simplex}
For $k\ge 2$ denote
\begin{eqnarray}\label{eq:int}
R(k)\coloneqq \frac{1}{k-1} \left( {\E} \left[  \max_{j \in
\{1,\ldots,k\}} g_j  \right] \right)^2=
\frac{k}{(2\pi)^{k/2}}\int_{-\infty}^\infty
xe^{-x^2/2}\left(\int_{-\infty}^x e^{-y^2/2}dy\right)^{k-1}dx.
\end{eqnarray}
Then for every integer $k\in \{2,4,5,\ldots\}$ we have $R(k)<
R(3)=\frac{9}{8\pi}$. Thus, if $v_1^k,\ldots,v_k^k$ are the vertices
of the regular simplex in $\R^{k-1}$ then for $k\ge 4$ we have
$$
\sum_{j=1}^k
\left\|\int_{P_j(v_1^k,\ldots,v_k^k)}xd\gamma_{k-1}(x)\right\|_2^2<\sum_{j=1}^3
\left\|\int_{P_j(v_1^3,v_2^3,v_3^3)\times
\R^{k-3}}xd\gamma_{k-1}(x)\right\|_2^2.
$$
\end{corollary}

\begin{proof} It follows from Corollary~\ref{coro:2,3} that
$R(3)=C(3)=\frac{9}{8\pi}$. We require a crude bound on $R(k)$. An
application of Stirling's formula shows that for $p\ge 2$ we have
$$
\left(\E\left[|g_1|^p\right]\right)^{1/p}=
\left(\frac{2^{p/2}}{\sqrt{\pi}}\Gamma\left(\frac{p+1}{2}\right)\right)^{1/p}\le
\sqrt{\frac{p}{2}}.
$$
Hence
$$
R(k)\le \frac{1}{k-1} \left( {\E} \left[  \max_{j \in
\{1,\ldots,k\}} |g_j|  \right] \right)^2\le
\frac{1}{k-1}\E\left[\left(\sum_{j=1}^k
|g_j|^p\right)^{1/p}\right]^2\le \frac{1}{k-1} \left(\sum_{j=1}^k
\E\left[|g_j|^p\right]\right)^{2/p}\le \frac{1}{k-1}\cdot
k^{2/p}\cdot \frac{p}{2}.
$$
Choosing $p=2\log k\ge 2\log 4>2$ we see that
\begin{eqnarray}\label{eq:crude}
R(k)\le \frac{e\log k}{k-1}.
\end{eqnarray}
The function $k\to \frac{\log k}{k-1}$ is decreasing on
$[4,\infty)$, and therefore a direct computation
using~\eqref{eq:crude} shows that $R(k)< \frac{9}{8\pi}$ for $k\ge
26$. For $k\le 25$ one can compute numerically (say, using Maple)
the integral in~\eqref{eq:int} and get the following values:
\begin{eqnarray*}
&&R(4)=0.3532045529,\ R(5)=0.3381215916,\ R(6)=0.3211623921,\
R(7)=0.3047310600,\\&& R(8)=0.2895196903, \ R(9)=0.2756580116 ,\
R(10)=0.2630844408,\ R(11)=0.2516780298,\\ &&R(12)=0.2413075184,\
R(13)=0.2318492693,\ R(14)=0.2231929784,\ R(15)=0.2152425349,\\
&& R(16)=0.2079150401,\ R(17)=0.2011392394,\ R(18)=0.1948538849,\
R(19)=0.1890062248,\\
&& R(20)=0.1835506894,\ R(21)=0.1784477705,\ R(22)=0.1736630840,\
R(23)=0.1691665868,\\
&&R(24)=0.1649319261,\ R(25)=0.1609358965.
\end{eqnarray*}
Since $R(3)=0.3580986219$ it follows that $R(k)<R(3)$ for every
integer $k\in [4,25]$ as well.
\end{proof}

We conjecture that $C(k)\le C(3)$ for every integer $k\ge 2$. For
future reference we end this section with the following alternative
characterization of $C(k)$:

\begin{lemma}\label{lem:alternative} We have the following identity:
\begin{eqnarray}\label{eq:alt} C(k) =  \sup\left\{
  \frac{ \left( {\E} \left[ \max_{j\in\{1,\ldots,k\}} g_j \right] \right)^2 }{\sum_{j=1}^k
    {\E}\left[g_j^2\right] }:\ (g_1,\ldots,g_k)\in \R^{k}\ \mathrm{mean\  zero\  Gaussian\  vector}\right\}.
    \end{eqnarray}
\end{lemma}

\begin{proof}
First we show that $C(k)$ is at most the right hand side
of~\eqref{eq:alt}. We know that there exists a partition
$A_1,\ldots,A_k$ of $\R^{k-1}$ such that if we write $z_i
\coloneqq\int_{A_i} x \ d \gamma_{k-1}(x)$ then
$A_j=P_j(z_1,\ldots,z_k)$for all $j\in \{1,\ldots, k\}$  and
$C(k)=\sum_{j=1}^k \|z_j\|_2^2$. Now,
 \begin{multline}\label{eq:pass to h}
 C(k) =   \sum_{j=1}^k \|z_j\|_2^2
=   \sum_{j=1}^k   \int_{P_j(z_1,\ldots,z_k)} \langle x ,z_j\rangle
\ d\gamma_{k-1}(x)
  =   \sum_{j=1}^k  \int_{P_j(z_1,\ldots,z_k)} \left( \max_{i\in \{1,\ldots,k\}}
    \langle x, z_i\rangle \right)  d\gamma_{k-1}(x) \\
     =  \int_{\R^{k-1}}  \left( \max_{i\in \{1,\ldots,k\}}
    \langle x, z_i\rangle \right)  d\gamma_{k-1}(x)
      = \E \left[ \max_{j\in \{1,\ldots,k\}} h_j \right],
\end{multline}
where in~\eqref{eq:pass to h} $h_1,\ldots,h_k$ are mean-zero
Gaussians with covariances ${\E} [h_i h_j]
 = \langle z_i, z_j\rangle$. 
  Thus
 $$ C(k) =
  \frac{ \left( \E \left[ \max_{j\in \{1,\ldots,k\}} \ h_j \right] \right)^2 }{\sum_{j=1}^k
    {\E}\left[h_j^2\right] }, $$
which implies the desired upper bound on $C(k)$.

%

\medskip
For the other direction fix a mean zero Gaussian vector
$(g_1,\ldots,g_k)\in \R^k$  and
let $v_1, v_2, \ldots, v_k\in \R^k$ be vectors such that
 ${\E} [g_i g_j ] = \langle v_i , v_j\rangle$ for all $i,j\in \{1,\ldots,k\}$.
For $i\in \{1,\ldots,k\}$ let $w_i \coloneqq
\int_{P_i(v_1,\ldots,v_k)} x \ d\gamma_{k-1}(x)$. Now,
 \begin{multline*}
\sqrt{ \left( \sum_{i=1}^k \| w_i \|_2^2\right)
   \left( \sum_{i=1}^k \| v_i \|_2^2\right) }  \geq
   \sum_{i=1}^k \langle w_i, v_i\rangle
    =  \sum_{i=1}^k   \int_{P_i(v_1,\ldots,v_k)} \langle x, v_i\rangle \ d\gamma_{k-1}(x)
\\ =    \sum_{i=1}^k  \int_{P_i(v_1,\ldots,v_k)} \left( \max_{j\in
\{1,\ldots,k\}}
    \langle x, v_j\rangle \right)  d\gamma_{k-1}(x)
     = \int_{\R^{k-1}}  \left( \max_{j\in \{1,\ldots, k\}}
    \langle x ,v_j\rangle \right)  d\gamma_{k-1}(x)
      =  {\rm E} \left[ \max_{j\in \{1,\ldots,k\}} \ g_j \right].
 \end{multline*}
Therefore,
$$ C(k) \geq \sum_{i=1}^k \| w_i \|^2 \geq
 \frac{   \left( {\E} \left[ \max_{j\in \{1,\ldots,k\}} \ g_j \right] \right)^2 }
  { \sum_{j=1}^k \| v_j \|_2^2  } =
   \frac{   \left( {\E} \left[ \max_{j\in \{1,\ldots,k\}} \ g_j \right] \right)^2 }
  { \sum_{j=1}^k {\E} \left[ g_j^2 \right] }.
   $$
   This completes the proof of~\eqref{eq:alt}.
\end{proof}

\subsection{Dictatorships vs. functions with small influences}

In what follows all functions are assumed to be measurable and we
use the notation $[k]\coloneqq\{1,\ldots,k\}$. In this section we
will associate to every function from $\{1,\ldots,k\}^n$ to
$$\Delta_k\coloneqq \left\{ x \in \R^k : \ x_i \geq 0 \ \wedge\
\forall \ i \in [k], \ \ \sum_{i=1}^k x_i \leq 1 \right\}$$ a
numerical parameter, or ``objective value". We will show that the
value of this parameter for functions which depend only on a single
coordinate (i.e. dictatorships) differs markedly from its value on
functions which do not depend significantly on any particular
coordinate (i.e. functions with small influences). This step is an
analog of the ``dictatorship test" which is prevalent in PCP based
hardness proofs.

We begin with some notation and preliminaries on Fourier-type
expansions. For any function $f: \R^n \to \Delta_k$ we write
$f=(f_1, f_2, \ldots, f_k)$ where $f_i : \R^n \to [0,1]$ and
$\sum_{i=1}^k f_i \leq 1$. With this notation we have
 $$ C(k) =\sup_{f : \R^{k-1} \to \Delta_k}  \sum_{i=1}^k  \left\| \int_{\R^{k-1}} x
 f_i(x) d\gamma_{k-1} (x) \right\|_2^2, $$
where $C(k)$ is as in Section~\ref{sec:geometry}. We have already
seen that the supremum above is actually attained and at the
supremum we have $\sum_{i=1}^k f_i = 1$. Also $C(k)$ remains the
same if the supremum is taken over functions over $\R^n$ with $n
\geq k-1$, i.e. for every $n \geq k-1$,
$$ C(k) = \sup_{f : \R^n \to \Delta_k}  \sum_{i=1}^k
 \left\| \int_{\R^n} x f_i(x) d\gamma (x) \right\|_2^2. $$


Let $(\Omega = [k], \mu)$ be a probability space, $\mu$ being the
uniform measure. Let $(\Omega^n, \mu^n)$ be the product space. We
will be analyzing functions $f: \Omega^n \to \Delta_k$ (and more
generally into $\R^k$). Fix a basis of orthonormal random variables
on $\Omega$ where one of them is the constant $1$, i.e.  $\{X_0,
X_1, \ldots, X_{k-1}\}$ where $\forall \ i, X_i : \Omega \to \R$,
$X_0 \equiv 1$ and $\E_{\omega \in \Omega} [ X_i (\omega) X_j
(\omega) ] = 0$ for $i\not=j$ and equal to $1$ if $i=j$. Then any
function $f: \Omega \to \R$ can be written as a linear combination
of the $X_i$'s.

In order to analyze functions $f: \Omega^n \to \R$, we let $\calX =
 (\calX_1, \calX_2, \ldots, \calX_n)$ be an ``ensemble" of random variables
 where for $1 \leq i \leq n$, $\calX_i = \{ X_{i,0}, X_{i,1}, \ldots, X_{i,k-1}\}$, and
 for every $i$,  $\{X_{i,j}\}_{j=0}^{k-1}$ are independent copies of  the $\{X_j\}_{j=0}^{k-1}$. Any
 $\sigma = (\sigma_1, \sigma_2, \ldots, \sigma_n) \in \{0,1,2,\ldots, k-1\}^n$ will be called a
 multi-index. We shall denote by $|\sigma|$ the number on non-zero
 entries in $\sigma$.
 Each multi-index defines a monomial $x_{\sigma}
  := \prod_{i \in [n], \sigma_i\not=0} x_{i, \sigma_i}$ on a set of $n(k-1)$ indeterminates $\{x_{ij} \ | \
   i \in [n], j \in \{1,2,\ldots, k-1\} \}$, and also a random variable
   $X_\sigma : \Omega^n \to \R$ as
    $$ X_\sigma (\omega) :=  \prod_{i=1}^n X_{i,\sigma_i} (\omega_i). $$
It is easy to see that the random variables $\{X_\sigma\}_\sigma$
form an orthonormal basis for the space of functions $f: \Omega^n
\to \R$. Thus, every such $f$ can be written uniquely as (the
``Fourier expansion")
 $$ f = \sum_{\sigma} \widehat{f}(\sigma) X_\sigma, \quad \widehat{f}(\sigma)
 \in \R. $$
We denote the corresponding multi-linear polynomial as $Q_f =
\sum_\sigma \widehat{f}(\sigma) x_\sigma$. One can think of $f$ as
the polynomial $Q_f$ applied to the ensemble $\calX$, i.e. $f =
Q_f(\calX)$. Of course, one can also apply $Q_f$ to any other
ensemble, and specifically to the Gaussian ensemble $\calG =
(\calG_1, \calG_2, \ldots, \calG_n)$ where $\calG_i = \{
G_{i,0}\equiv 1, G_{i,1},\ldots, G_{i, k-1}\}$ and $G_{i,j}, i \in
[n], 1 \leq j \leq k-1$ are i.i.d. standard Gaussians. Define the
influence of the $i$'th variable on $f$ as
$$
\infl_i(F)\coloneqq \sum_{\sigma_i\neq 0} \widehat f(\sigma)^2.
$$
Roughly speaking, the results of~\cite{Rotar79,MOO}
 say that if $f: \Omega^n \to [0,1]$ is a
function with all low influences, then $f= Q_f(\calX)$ and
$Q_f(\calG)$ are almost identically distributed, and in particular,
the values of $Q_f(\calG)$ are essentially contained in
 $[0,1]$. Note that $Q_f(\calG)$ is a random variable on
the probability space $(\R^{n (k-1)}, \gamma_{n(k-1)})$.



Consider functions $f: \Omega^n \to \Delta_k$. We write $f= (f_1,
f_2, \ldots, f_k)$ where $f : \Omega^n \to [0,1]$ with $\sum_{i=1}^k
f_i \leq 1$. Each $f_i$ has a unique representation (along with the
corresponding multi-linear polynomial)
 $$f_i = \sum_\sigma \widehat{f_i}(\sigma) X_\sigma,  \quad \quad Q_i := Q_{f_i} =
  \sum_{\sigma} \widehat{f_i}(\sigma) x_\sigma.  $$

We shall define an objective function ${\rm OBJ}(f)$ that is a
positive semi-definite quadratic form on the table of values of $f$.
Then we analyze the value of this objective function when $f$ is a
``dictatorship" versus when $f$ has all low influences.

\subsubsection*{The objective value} For a function $f: \Omega^n \to
\Delta_k$ (or more generally, $f: \Omega^n \to \R^k$) define
\begin{equation}
 \mbox{OBJ} (f)   :=  \sum_{i=1}^k \sum_{\sigma:\  |\sigma|=1} \widehat{f_i}( \sigma)^2.
\end{equation}
In words, $\mbox{OBJ}(f)$ is the total ``Fourier mass" of all
functions $\{f_i\}_{i=1}^k$ at level $1$. Note that there are
$n(k-1)$ multi-indices $\sigma$ such that $|\sigma|=1$.

\subsubsection*{The objective value for dictatorships} For $\ell \in
[n]$ we define a dictatorship function $f^{dict, \ell}: \Omega^n \to
\Delta_k$ as follows. The range of the function is limited to only
$k$ points in $\Delta_k$, namely the points $\{e_1, e_2, \ldots,
e_k\}$ where $e_i$ is a vector with $i^{th}$ coordinate $1$ and all
other coordinates zero.

\begin{equation}
  f^{dict, \ell} (\omega)  :=  e_i \  \mbox{if}  \ \omega_\ell = i.
\end{equation}
In other words, when one writes
 $ f^{dict, \ell}  = (f_1, f_2, \ldots, f_k)$, $f_i$ is  $\{0,1\}$-valued and
 $f_i(\omega) =1$ iff $\omega_\ell = i$. It is easy to see that the Fourier expansion of $f_i$ is
 \begin{eqnarray}\label{eq:expand dict}
 f_i (\omega) = \frac{1}{k} \sum_{\sigma:\  \sigma_j=0 \ \forall j \not= \ell}
  X_{\sigma_\ell}(i) \    X_\sigma (\omega).
   \end{eqnarray}
Indeed, the right hand side of~\eqref{eq:expand dict} equals
\begin{eqnarray*}
 \frac{1}{k} \sum_{0 \leq \sigma_\ell \leq k-1}
  X_{\sigma_\ell}(i) \    X_{\sigma_\ell} (\omega_\ell) =
  \left\{ \begin{array}{l}
           1 \ \  \mbox{if} \ \omega_\ell = i, \\
           0 \ \  \mbox{otherwise.}
           \end{array} \right.
\end{eqnarray*}
The Fourier mass of $f_i^{dict, \ell}$ at level $1$ equals
$$ \sum_{1 \leq \sigma_\ell \leq k-1} \left( \frac{X_{\sigma_\ell}(i)}{k} \right)^2
 = - \left( \frac{X_{0}(i)}{k} \right)^2 + \sum_{0 \leq \sigma_\ell \leq k-1} \left( \frac{X_{\sigma_\ell}(i)}{k} \right)^2  = -\frac{1}{k^2} + \frac{k}{k^2} = \frac{k-1}{k^2}.$$
Summing the Fourier mass of all $f_i^{dict, \ell}$'s at level $1$,
we get
\begin{equation}
  \mbox{OBJ} (f^{dict, \ell}) =  1-\frac{1}{k}.   \label{eqn:dict-completeness}
\end{equation}

\subsubsection*{The objective value for functions with low
influences} For $f:\Omega^n\to \R$, $j\in [n]$ and $m\in \mathbb N$
denote
$$
\infl_j^{\le m}(f)\coloneqq \sum_{\substack{|\sigma|\le
m\\\sigma_j\neq 0}}\widehat f(\sigma)^2.
$$
For every $\eta>0$ we will use the smoothing operator:
$$
T_\eta f = \sum_{\sigma} \eta^{|\sigma|}\widehat{f}(\sigma)
X_\sigma.
$$
The following theorem is the key analytic fact used in our UGC
hardness result:

\begin{theorem}\label{thm:dict-soundness}
 For every $\eps > 0$, there exists $\tau > 0$  so that the following holds: for any function
 $f : \Omega^n\to \Delta_k$ such that
 $$\forall\  i \in [k],\ \forall\  j \in [n],  \quad \infl_j^{\leq \log (1/\tau)}(f_i) \leq \tau
 $$
 we have,
 $$  {\rm OBJ}(f) \leq C(k)+\eps. $$
\end{theorem}

\begin{proof} Let $\delta, \eta > 0$ be sufficiently small constants to be chosen later.
Let   $Q_i= Q_{f_i}$ be the multi-linear polynomial associated with
$f_i$. Recall that $Q_i$ is a multi-linear polynomial in  $n(k-1)$
indeterminates
 $\{ x_{j\ell} \ | \ j \in [n], \ell \in [k-1]\}$.  Moreover $f_i = Q_i(\calX)$
has range $[0,1]$ and $\sum_{i=1}^k f_i \leq 1$.


Let $R_i = (T_{1-\delta} Q_i) (\calX) $ and $S_i = (T_{1-\delta}
Q_i) (\calG)$ (the smoothening operator $T_{1-\delta}$ helps  us
meet some technical pre-conditions before applying the invariance
principle on~\cite{MOO}).
 Note that $R_i$ has range $[0,1]$ and $S_i$ has range $\R$. It will follow however from~\cite{MOO}
 that $S_i$ is with high probability in $[0,1]$. First we relate
${\rm OBJ}(f)$ to the functions $S_i$ which will, up to truncation,
induce a partition of $\R^{n(k-1)}$, which in turn will give the
bound in terms of $C(k)$.



\begin{eqnarray}
(1-\delta)^2 \cdot \mbox{OBJ}(f)
   & = & (1-\delta)^2 \sum_{i=1}^k \sum_{\sigma: |\sigma|=1} \widehat{f_i}(\sigma)^2 \nonumber \\
   & = & (1-\delta)^2 \sum_{i=1}^k \sum_{j=1}^n \sum_{\ell=1}^{k-1}
    \left| \int_{\R^{n(k-1)}}
  x_{j\ell}  \ Q_i (x) d\gamma_{n(k-1)} (x)  \right|^2 \nonumber \\
  & = & (1-\delta)^2 \sum_{i=1}^k \left\| \int_{\R^{n(k-1)}}
  x \ Q_i (x) d\gamma_{n(k-1)} (x)  \right\|_2^2 \nonumber \\
   & = &  \sum_{i=1}^k \left\| \int_{\R^{n(k-1)}}
  x \ (T_{1-\delta}Q_i) (x) d\gamma_{n(k-1)} (x)  \right\|_2^2 \nonumber \\
  & = & \sum_{i=1}^k \left\| \int_{\R^{n(k-1)}}
  x \ S_i (x) d\gamma_{n(k-1)} (x)  \right\|_2^2.   \label{eqn:estimate-Rn}
\end{eqnarray}
We bound the last term by $C(k)+ o(1)$. For any real-valued function
$h$ on $\R^{n (k-1)}$, let
   $$ \chop(h) (x) :=   \left\{ \begin{array}{ll}
                         0 & \mbox{if} \ h(x) < 0, \\
                         h(x) & \mbox{if} \ h(x) \in [0,1], \\
                         1 & \mbox{if} \ h(x) > 1.
                           \end{array}  \right.  $$
For every subset $I \subseteq [k]$, let $Q_I := \sum_{i \in I} Q_i$.
Since every $Q_i$ has small low-degree influence, so does every
$Q_I$. Let
 $R_I \coloneqq \sum_{i \in I} (T_{1-\delta} Q_i) (\calX) $, and
 $S_I \coloneqq \sum_{i \in I} (T_{1-\delta} Q_i) (\calG)$. Note that $R_{ \{i\} }=R_i$
 and $S_{ \{ i \} }=S_i$. Applying Theorem 3.20 in~\cite{MOO} to the polynomial $Q_I$, it
follows that (provided $\tau$ is sufficiently small compared to
$\delta$ and $\eta$),
 \begin{eqnarray} \label{error-term-1}
 \left\| S_I  - \chop(S_I) \right\|_2^2 =  \int_{\R^{n(k-1)}}
 \left|S_I(x) - \chop(S_I)(x) \right|^2 d\gamma_{n(k-1)} (x) \leq \eta.
 \end{eqnarray}


%
%

The functions $\chop(S_i)$ are almost what we want except that they
might not sum up to at most $1$. So further define
\begin{equation}
 S^*_i(x) := \left\{ \begin{array}{ll}
                     \chop(S_i) (x)  & \mbox{if} \ \sum_{i=1}^k \chop(S_i) (x) \leq 1, \\ \\
                     \frac{\chop(S_i)(x)}{(\sum_{i=1}^k  \chop(S_i) (x))}
                     & \mbox{if} \  \sum_{i=1}^k \chop(S_i) (x) > 1.
                     \end{array} \right. \nonumber
\end{equation}
Clearly, $S^*_i$ have range $[0,1]$ and  $\sum_{i=1}^k S^*_i \leq
1$. Observe that the following holds point-wise:


$$ 0 \leq \chop(S_i) - S_i^* \leq \sum_{j=1}^k \left( \chop(S_j) - S_j^* \right)
 \leq \max \left(0, \sum_{j=1}^k \chop(S_j)-1\right)  \leq
  \sum_{I \subseteq [k]}
   \left|S_I - \chop(S_I)\right|, $$
where the last inequality holds since for every $x$, by defining $I
=  I(x) =
 \{ j \ | \ S_j(x) \geq 0 \}$,
$$  \sum_{j=1}^k \chop(S_j)(x)-1 =
 \sum_{j \in I} \chop(S_j)(x) -1 \leq \sum_{j \in I} S_j(x) -1 \leq
   \left| S_I(x) - \chop(S_I) (x) \right|.$$
It follows that
$$\left\|\chop(S_i) - S_i^*\right\|_2 \leq \sum_{I\subseteq [k]} \left\|S_I -
\chop(S_I) \right\|_2
 \leq 2^k \sqrt{\eta},
 $$
 where we used (\ref{error-term-1}).  Finally,
\begin{equation}
\left\| S_i - S^*_i \right\|_2 \leq \left\|S_i - \chop(S_i)
\right\|_2 + \left\|\chop(S_i) - S^*_i\right\|_2 \leq
  (2^k+1)\sqrt{\eta}.   \label{eqn:error-term-2}
\end{equation}   Now write
\begin{eqnarray}\label{eq:two integrals} \int_{\R^{n(k-1)}}
  x \ S_i (x) d\gamma_{n(k-1)} (x)  =  \int_{\R^{n(k-1)}}
  x \ S^*_i (x) d\gamma_{n(k-1)} (x) +  \int_{\R^{n(k-1)}}
  x \ (S_i (x) - S^*_i
  (x)) d\gamma_{n(k-1)} (x). \end{eqnarray}
The norm of second integral is bounded by $(2^k+1)\sqrt{\eta}$ using
 (\ref{eqn:error-term-2}) and Lemma
\ref{lemma:trivial-bound} below.  Since $\| S^*_i \|_2 \leq 1$, the
norm of first integral is bounded by $1$, and thus
$$ \left\| \int_{\R^{n(k-1)}}
  x \ S_i (x) d\gamma_{n(k-1)} (x) \right\|_2^2  \leq  \left\|  \int_{\R^{n(k-1)}}
  x \ S^*_i (x) d\gamma_{n(k-1)} (x)  \right\|_2^2  + 2 (2^k+1) \sqrt{\eta} + (2^k+1)^2
  \eta.
   $$

%
%
%
%
%

Returning to the estimation in Equation (\ref{eqn:estimate-Rn}) and
noting that $\sum_{i=1}^k S^*_i \leq 1$,
\begin{eqnarray*}
\sum_{i=1}^k \left\| \int_{\R^{n(k-1)}}
  x \ S_i (x) d\gamma_{n(k-1)} (x)  \right\|_2^2 & \leq &  \sum_{i=1}^k \left(
  \left\|  \int_{\R^{n(k-1)}}
  x \ S^*_i (x) d\gamma_{n(k-1)} (x)  \right\|_2^2  + 2 (2^k+1)^2 \sqrt{\eta} \right)  \\
  & \leq & \sup_{f: \R^{n(k-1)} \to \Delta_k} \left( \sum_{i=1}^k \left\|  \int_{\R^{n(k-1)}}
  x \ f_i (x) d\gamma_{n(k-1)} (x)  \right\|_2^2   \right)  + 2(2^k+1)^3\sqrt{\eta} \\
  & = & C(k) +  2(2^k+1)^3\sqrt{\eta}.
\end{eqnarray*}
It follows that $\mbox{OBJ}(f) \leq \frac{C(k) + 2(2^k+1)^3
\sqrt{\eta}}{1-\delta^2}
 \leq C(k) + \eps $, provided that $\eta$ and $\delta$ are small
 enough.
\end{proof}

\begin{lemma}\label{lemma:trivial-bound}
 Let $g \in L_2(\R^n, \gamma_n)$. Then
 $$ \left\| \int_{\R^n} x \ g(x) d\gamma_n (x) \right\|_2  \leq  \|g \|_2. $$
\end{lemma}
\begin{proof} Note that the square of the left hand side  equals
 $$ \sum_{i=1}^n \left| \int_{\R^n} x_i \ g(x)  d\gamma_n (x) \right|^2
  =  \sum_{i=1}^n  \langle x_i, g \rangle^2.  $$
Since $x_i \in  L_2(\R^n, \gamma_n)$ are an orthonormal set of
functions, the sum of squares of projections of $g$ onto them is at
most the squared norm of $g$.
\end{proof}

\subsubsection*{The intended hardness factor}

As we show next, the dictatorship test can be translated (in a more
or less standard way by now) into a UGC-hardness result. The
hardness factor (as usual) turns out to be
 the ratio of the objective value when the function is a
dictatorship versus the function has all low influences, i.e.
$$\frac{1-1/k}{C(k)+o(1)}
 = \frac{1-1/k}{C(k)} - o(1).
 $$


\subsection{The reduction from unique games to kernel clustering}

Given a Unique Games Instance $\calL(G(V,W,E), [n], \{\pi_{vw}: [n]
\to [n]  \}_{(v,w)\in E})$, we construct an instance  of the
clustering problem. We first reformulate the kernel clustering
problem for the ease of presentation.

\subsubsection*{Reformulation of the problem} Given an instance of
the kernel clustering problem $(A = (a_{st}), B = (b_{ij}))$ where
$A$ and $B$ are $N \times N$ and $k \times k$ PSD matrices
respectively, we note that \begin{eqnarray}
 \max_{\atop{ \sigma: [N] \to [k]}} \sum_{s,t} a_{st}\  b_{\sigma(s), \sigma(t)}
  & = & \max_{\atop{ F: [N] \to \Delta_k}} \sum_{s,t} a_{st} \sum_{ij} b_{ij}
   F(s)_i F(t)_j  \label{eqn:justification-1} \\
  & = & \max_{\atop{ F: [N] \to \Delta_k}} \sum_{i,j} b_{ij}
    \sum_{s,t} a_{st} F_i(s) F_j(t) \label{eqn:justification-2} \\
  & = &  \max_{\atop{ F: [N] \to \Delta_k}} \sum_{i,j} b_{ij}
     Q_A( F_i, F_j)   \label{eqn:justification-3}
\end{eqnarray}
where on line (\ref{eqn:justification-1}), instead of choosing a
label $\sigma(s) \in [k]$, we allow a distribution over the $k$
labels $F(s) \in \Delta_k$. The equality follows since any such
probabilistic labeling $F$ yields a labeling $\sigma$ with the same
expected objective value by picking, for every $s \in [N]$,  a label
$i$ with probability $F(s)_i$. On line (\ref{eqn:justification-2})
we interchanged the order of summation and interpreted the $i^{th}$
co-ordinate of $F(s)$ (i.e. $F(s)_i$)  as the value of a function
$F_i: [N] \to [0,1]$ at index $s$ (i.e. $F_i(s)$). Thus $F = (F_1,
F_2, \ldots, F_k)$. On line (\ref{eqn:justification-3})  we rewrote
 $\sum_{s,t} a_{st} F_i(s) F_j(t)$ as a PSD quadratic form $Q_A(F_i, F_j)$ on the tables of
 values of functions $F_i$ and $F_j$.

\medskip

This enables us to reformulate the clustering problem as: Given a
PSD matrix B, and a PSD quadratic form $Q(\cdot ,\cdot )$ on $\R^N
\times \R^N$, find
 $F: [N] \to \Delta_k$,  $F= (F_1, F_2,\ldots, F_k)$, so as to maximize
 $\sum_{ij} b_{ij} Q(F_i, F_j)$.

\subsubsection*{The clustering problem instance} Given a Unique
Games instance $$\L\left(G(V,W,E), [n], \{\pi_{vw}: [n] \to [n]
\}_{(v,w)\in E}\right),$$ the clustering  problem is to find $F: W
\times \Omega^n \to \Delta_k$ so as to maximize $\sum_{i=1}^k Q(F_i,
F_i)$ where $Q$ is a suitably defined PSD quadratic form. Thus the
matrix $B$ is the $k \times k$ identity matrix. For notational
convenience, we let
   $$ F_w :=  F(w, \cdot),  \quad\quad F_w : \Omega^n \to \Delta.$$
Also, for every $v \in V$, we let
$$  F_v \coloneqq  \E_{(v,w) \in E} \left[ F_w  \circ \pi_{vw} \right], \quad\quad\quad
  F_v : \Omega^n \to \Delta.  $$
We used the following notation: for any function $g : \Omega^n \to
\Delta_k$ and $\pi : [n] \to [n]$,  $g \circ \pi: \Omega^n \to
\Delta_k$ denotes a function
 $$ (g \circ \pi) (\omega) :=  g ( \omega_{\pi(1)}, \omega_{\pi(2)}, \ldots, \omega_{\pi(n)} ). $$
As usual, we denote $F_w = (F_{w,1}, F_{w,2}, \ldots, F_{w,k})$
where each $F_{w,i}$ has range $[0,1]$ and $\sum_{i=1}^k F_{w,i}
\leq 1$. Similarly, $F_v = (F_{v,1}, F_{v,2}, \ldots, F_{v,k})$. Now
we are ready to define the clustering problem instance.

\medskip\noindent {\bf Clustering instance:} The goal is to find
$F:  W \times \Omega^n \to \Delta_k$ so as to maximize:
\begin{eqnarray}\label{eqn:clustering-obj}
  \max_{F: W \times \Omega^n \to \Delta_k} \ \
   {\E}_{v \in V} \left[ \mbox{OBJ} (F_v) \right] =
    \max_{F: W \times \Omega^n \to \Delta_k} \ \
   \sum_{i=1}^k \ \E_{v \in V} \left[ \sum_{\sigma: |\sigma|=1} \widehat{F}_{v,i} (\sigma)^2
   \right].
\end{eqnarray}

%

\subsubsection*{Completeness.} We will show that if the Unique
Games instance has an almost satisfying labeling, then the objective
value of the clustering problem is $(1-o(1))\cdot(1-1/k)$. So, let
$\rho: V \cup W \to [n]$ be the labeling, such that for at least
$1-\eps$ fraction of the vertices $v \in V$ (call such $v$ good) we
have
 $$  \pi_{vw} ( \rho(w) ) = \rho(v) \ \ \forall \ \ (v,w) \in E. $$
Define $F: W \times \Omega^n \to \Delta_k$ as follows: for every $w
\in W$,
 $F_w: \Omega^n \to \Delta_k$
  equals the dictatorship for $\rho(w) \in [n]$, i.e.
   $$ F_w :=   f^{dict, \rho(w)}.$$

\begin{lemma} $f^{dict, j} \circ \pi =  f^{dict, \pi(j)}$.
\end{lemma}
\begin{proof}
$f^{dict, \pi(j)}(\omega)$ equals $e_\ell$ if $\omega_{\pi(j)}=
\ell$.  On the other hand $$(f^{dict,j}\circ \pi)(\omega) = f^{dict,
j} (\omega_{\pi(1)}, \omega_{\pi(2)}, \ldots, \omega_{\pi(n)}),$$
which equals $e_\ell$ since $\omega_{\pi(j)} = \ell$.
\end{proof}

\begin{lemma} For a good $v\in V$,   $F_v = f^{dict, \rho(v)}$.
\end{lemma}
\begin{proof} For a good $v$, $\pi_{vw}(\rho(w))=\rho(v)$ for every $(v,w)\in E$. Thus
\begin{multline*}
F_v  = \E_{(v,w)\in E}
   \Big[ \  F_w \circ \pi_{vw} \Big]
     =
    \E_{ (v,w)\in E}
   \Big[ \  f^{dict, \rho(w)} \circ \pi_{vw} \Big]
    \\=  \E_{ (v,w)\in E}
   \Big[ \  f^{dict, \pi_{vw}(\rho(w))}  \Big]
    =  \E_{ (v,w)\in E}
   \Big[ \  f^{dict, \rho(v)}   \Big] =  f^{dict, \rho(v)}
\end{multline*}
\end{proof}

Thus the contribution of $v$ in  (\ref{eqn:clustering-obj}) is
$\mbox{OBJ}(f^{dict, \rho(v)}) = 1-1/k$
as observed in Equation (\ref{eqn:dict-completeness}). Since
$1-\eps$ fraction of $v \in V$ are good, (\ref{eqn:clustering-obj})
is at least $(1-\eps)\cdot (1-1/k).$

\subsubsection*{Soundness} Suppose on the contrary that the value
of (\ref{eqn:clustering-obj}) is at least $C(k) + 2\eps$. We will
prove that the Unique Games instance must have a labeling that
satisfies at least $\frac{\eps \tau^2}{4k\log(1/\tau)}$ fraction of
its edges, reaching a contradiction, provided its soundness is
chosen to be lower to begin with.

\medskip

We define a labeling as follows. First we define a not-too-large set
of labels $L(w) \subseteq [n]$ for every $w \in W$. Let $\tau$ be as
in Theorem \ref{thm:dict-soundness}.
 $$ L(w) := \left\{ j \in [n] \ | \ \exists \ i \in [k], \ \
  \infl_j^{\leq \log(1/\tau)} (F_{w, i}) \geq \tau/2 \right\} $$
Clearly, $|L(w)| \leq \frac{2k\log(1/\tau)}{\tau}$ since each
$F_{w,i}$ has range $[0,1]$ and therefore the sum of all
degree-$\log(1/\tau)$ influences is at most $\log(1/\tau)$.

\medskip

Now assume that the value of (\ref{eqn:clustering-obj}) is at least
$C(k) + 2\eps$. By an averaging argument,  for at least $\eps$
fraction of $v \in V$ (call such $v$ nice),
 $\mbox{OBJ}(F_v) \geq C(k)+\eps$. Applying Theorem \ref{thm:dict-soundness}, we conclude that
 there exists $i_0 \in [k], j_0 \in [n]$ such that $\infl_{j_0}^{\leq \log(1/\tau)} (F_{v,i_0})
  \geq \tau$. Observe that
\begin{eqnarray*}
\tau & \leq &  \infl_{j_0}^{\leq \log(1/\tau)} (F_{v,i_0}) \\
& = & \infl_{j_0}^{\leq \log(1/\tau)}\Big( \mbox{E}_{(v,w)\in E}
  \left[ F_{w,i_0} \circ \pi_{vw} \right] \Big)   \\
& \leq &  \mbox{E}_{(v,w) \in E} \left[  \infl_{j_0}^{\leq
\log(1/\tau)}\left(
   F_{w,i_0} \circ \pi_{vw}  \right) \right]  \quad\quad\quad\quad \mbox{Using \ Lemma \
   \ref{lemma:expected-infl}}\ \mathrm{below}  \\
& = &  \E_{(v,w) \in E} \left[ \infl_{\pi_{vw}^{-1}(j_0)}^{\leq
\log(1/\tau)}\left(
   F_{w,i_0} \right) \right] \quad\quad\quad\quad\quad\quad\ \ \! \mbox{Using \ Lemma \ \ref{lemma:permutation-infl}\ below}
\end{eqnarray*}
This implies that for at least $\frac{\tau}{2}$ fraction of $w$ such
that $(v,w) \in E$, we have $\frac{\tau}{2} \leq
\infl_{\pi_{vw}^{-1}(j_0)}^{\leq \log(1/\tau)}\left(
   F_{w,i_0} \right)$. Thus $\pi_{vw}^{-1}(j_0) \in L(w)$ by the definition of $L(w)$. Define $j_0$ to be
   the label of $v$. Finally, for every $w \in W$, select a random label from $L(w)$ (or an arbitrary label
   if $L(w) =\emptyset$). Noting that $\eps$ fraction of $v \in V$ are nice, and
   $|L(w)| \leq \frac{2k\log(1/\tau)}{\tau}$, it follows that the labeling satisfies $\ \eps \cdot \frac{\tau}{2} \cdot
   \frac{1}{2k\tau^{-1}\log(1/\tau)} = \frac{\eps \tau^2}{ 4k\log(1/\tau)}$ fraction of the edges of the Unique Games instance.

\begin{lemma}\label{lemma:expected-infl}
 Suppose $\calC$ is a class of functions $g : \Omega^n \to \R$ and
$h := \E_{g \in \calC} [g]$. Then for any $j \in [n]$ and integer
$d$,
 $$ \infl_j(h) \leq {\rm E}_{g \in \calC} \left[ \infl_j(g) \right], \quad\quad\quad
\infl_j^{\leq d}(h) \leq {\rm E}_{g \in \calC} \left[ \infl_j^{\leq
d}(g) \right].$$
\end{lemma}
\begin{proof} We prove the first inequality, the second is similar by restricting summations to
multi-indices $|\sigma| \leq d$.
$$ \infl_j (h) := \sum_{\sigma: \sigma_j \not= 0 } \widehat{h}(\sigma)^2 =
\sum_{\sigma: \sigma_j \not= 0}
 \left( \E_{g \in \calC} \left[ \widehat{g}(\sigma) \right] \right)^2
 \leq \sum_{\sigma: \sigma_j \not= 0} \E_{g \in \calC} \left[ \widehat{g}(\sigma)^2 \right]
=  \E_{g\in \calC} \left[ \infl_j (g) \right]. $$
\end{proof}

\begin{lemma} Suppose $g : \Omega^n \to \R$,
 $\pi : [n] \to [n]$ and let $\sigma$ be a multi-index.  Then
 $$ \widehat{g \circ \pi} (\sigma)  = \widehat{g}( \pi^{-1}(\sigma) ).$$
\end{lemma}
\begin{proof}
The proof is a straightforward computation which we omit.
\end{proof}

\begin{lemma} \label{lemma:permutation-infl}  Suppose $g : \Omega^n \to \R$,
 $\pi : [n] \to [n]$ and
$ j \in [n]$. Then
 $$ \infl_j (g \circ \pi) = \infl_{\pi^{-1}(j)} (g), \quad\quad\quad
\infl_j^{\leq d} (g \circ \pi) = \infl_{\pi^{-1}(j)}^{\leq d} (g).
$$
\end{lemma}
\begin{proof}
We prove the first equality, the second is similar by restricting
summations to multi-indices $|\sigma| \leq d$.
$$ \infl_j (g \circ \pi) := \sum_{\sigma: \sigma_j \not= 0} \widehat{g \circ \pi}(\sigma)^2
 = \sum_{\sigma: \sigma_j \not= 0}
 \widehat{g}(\pi^{-1}(\sigma))^2
 = \sum_{\overline{\sigma}:  \overline{\sigma}_{\pi^{-1}(j)}\not= 0} \widehat{g}(\overline{\sigma})^2 =
  \infl_{\pi^{-1}(j)}(g). $$
\end{proof}


\subsection*{Acknowledgements}
We thank Alex Smola for bringing the problem of approximation
algorithms for kernel clustering to our attention and for
encouraging us to publish our results.


\bibliographystyle{abbrv}

\bibliography{smola}

\end{document}